\documentclass[a4paper,11pt]{article}

\pdfoutput=1

\usepackage{graphicx}
\usepackage{color}
\usepackage{latexsym,amssymb,amsmath}
\usepackage[mathscr]{euscript}
\usepackage{graphicx}
\usepackage{chemarr}
\usepackage{functan}
\usepackage{subfig}
\usepackage{algpseudocode,algorithm,algorithmicx}
\algnewcommand{\Let}[2]{#1 $\gets$ #2}
\algnewcommand{\SLet}[2]{\State\Let{#1}{#2}}
\algnewcommand{\IIf}[1]{\State\algorithmicif\ #1\ \algorithmicthen}
\algnewcommand{\EndIIf}{\unskip\ \algorithmicend\ \algorithmicif}

\usepackage{booktabs}
\usepackage{xspace}

\newcommand{\euler}{\ensuremath{\mbox{euler}}\xspace}
\newcommand{\beuler}{\ensuremath{\mbox{beuler}}\xspace}
\newcommand{\rkfourfive}{\ensuremath{\mbox{rk45}}\xspace}

\newcommand{\N}{\mathbb N}

\newcommand{\heps}{\hat{\varepsilon}}

\newcommand{\teps}{\tilde{\varepsilon}}
\newcommand{\sig}{\ensuremath{\mathit{Sig}}}

\newcommand{\prob}{\ensuremath{\mathit{p}}}
\newcommand{\nextprob}{\ensuremath{\mathit{p_2}}}

\newcommand{\rtol}{\mbox{RTOL}}
\newcommand{\atol}{\mbox{ATOL}}

\newcommand{\lsrtol}{\mathit{rtol}}    
\newcommand{\lsatol}{\mathit{atol}}    

\newcommand{\zvec}{\mathbf{0}}

\graphicspath {{examples/}}

\begin{document}

\begin{center}
{\large\bfseries Approximate Numerical Integration of the Chemical Master Equation for Stochastic Reaction Networks}

\vspace*{1.5ex}

Linar Mikeev and Werner Sandmann

\vspace*{1.5ex}

Department of Computer Science, Saarland University, Saarbr\"ucken, Germany\\
linar.mikeev@uni-saarland.de, werner.sandmann@uni-saarland.de
\end{center}

\begin{abstract}
Numerical solution of the chemical master equation for stochastic reaction
networks typically suffers from the state space explosion problem due to the
curse of dimensionality and from stiffness due to multiple time scales.
The dimension of the state space equals the number of molecular species
involved in the reaction network and the size of the system of differential
equations equals the number of states in the corresponding continuous-time
Markov chain,
which is usually enormously huge and often even infinite.
Thus, efficient numerical solution approaches must be able to handle huge,
possibly infinite and stiff systems of differential equations efficiently.
We present an approximate numerical integration approach that combines a
dynamical state space truncation procedure with efficient numerical integration
schemes for systems of ordinary differential equations including adaptive step
size selection based on local error estimates.
The efficiency and accuracy is demonstrated by numerical examples.
\end{abstract}



\section{Introduction}
\label{sec:intro}
The chemical master equation (CME) describes the random dynamics of many
stochastic reaction networks in physics, biology, and chemistry, amongst
other sciences.
More specifically, many biochemical reaction networks 
can be appropriately modeled by multivariate continuous-time Markov chains
(CTMCs), also referred to as Markov jump processes, where at any time the
system state is represented by a vector of the numbers of each molecular
species present in the reaction network and the CME is a system of differential
equations whose solution, given an initial probability distribution, provides
the transient (time-dependent) state probabilities.

Exact analytical solutions of the CME are only available for very small
reaction networks or special cases such as, e.g., monomolecular reactions
or reversible bimolecular reactions \cite{jahnke-huisinga07,laurenzi00}.
Therefore, in general computational approaches are required.
However, because reaction rates typically differ by several orders of magnitude,
the system dynamics possess multiple time scales and the corresponding
equations are stiff. Furthermore, the size of the state space typically
increases exponentially with the model dimensionality, that is, with the
number of molecular species in the reaction network. This effect is often
referred to as the curse of dimensionality or state space explosion.
Stochastic simulation of the reaction network and the numerical
solution of the CME are two common complementary approaches
to analyze stochastic reaction networks governed by the CME, both of
which must be properly designed to cope with huge, potentially
infinite multidimensional state spaces, and stiffness.

Stochastic simulation does not solve the CME directly but imitates the
reaction network dynamics by generating trajectories (sample paths) of
the underlying CTMC, where a stochastically exact imitation is in principle
straightforward \cite{bortz-etal75,gillespie76,gillespie77}.
Several mathematically equivalent implementations
have been developed
\cite{cao-li-petzold04,gibson-bruck00,li-petzold06,mccollum-etal06,sandmann-cbac08}.
Stochastic simulation does not suffer from state space explosion, because the
state space needs not be explicitly enumerated, but in particular stochastic
simulation of stiff systems becomes exceedingly slow and inefficient, because
simulating all successive reactions in order to generate trajectories of the
underlying CTMC advances only in extremely small time steps.
In order to tackle this
problem approximate stochastic simulation techniques for accelerated
trajectory generation have been developed, in particular various
tau-leaping methods
\cite{anderson08-postleap-tau,cao-etal06,cao-etal07-adaptive,gillespie01,rathinam-samad07,rathinam-etal03-stiffness,sandmann-wsc09,tian-burrage04-binomialtau,xu-cai08}
where, instead of simulating every reaction, at any time $t$ a time
step size $\tau$ is determined by which the simulation is advanced.

A general problem of stochastic simulation, however, remains also with approximate techniques: stochastic
simulation constitutes an algorithmic statistical estimation procedure that
tends to be computationally expensive and only provides estimates whose
reliability and statistical accuracy in terms of relative errors or confidence interval
half widths depend on the variance of the corresponding simulation estimator.
Estimating the whole probability distribution by stochastic simulation is 
enormously time consuming. Therefore, often only statistical estimates of
the expected numbers of molecules are considered, which requires less effort,
but in any case, when estimating expectations, probabilities, or any other relevant
system property, many stochastically independent and identically distributed
trajectories must be generated in order to achieve a reasonable statistical
accuracy \cite{sandmann09}. Hence, stochastic simulation is inherently costly.
In many application domains it is sometimes even referred to as a method of last
resort.

Several hybrid approximation approaches combine stochastic simulation with
deterministic numerical computations. One way to do this is by distinguishing
between low molecular counts, where the evolution is described by the CME and
handled by stochastic simulation, and high molecular counts, where an
approximation of the CME by the continuous-state Fokker-Planck equation is
viable \cite{hellander-loetstedt07,safta-etal15}.
Another way is to consider time scale separations where parts of the system
(states or reactions) are classified as either slow or fast and the different
parts are handled by different stochastic simulation approaches and numerical
solution techniques
\cite{burrage-etal04,e-etal07-nested_ssa,harris-clancy06,haseltine-rawlings02,macnamara-etal08a,macnamara-etal08b,rao-arkin03,salis-kaznessis05,vasudeva-bhalla04},
or even without resorting to stochastic simulation for any part
\cite{busch-sandmann-wolf06,sandmann-wolf08}.
As a major drawback, however, for time scale separation methods it is in
general hard to define what is slow or fast. In fact, many systems possess
multiple time scales rather than only two and a clear separation of time
scales is impossible.
%
%

Some numerical approximation approaches tackle the state space explosion
problem by restricting the analysis of the model to certain subsets of
states 
where the truncated part of the state space has only a sufficiently small
probability. For instance, finite state projection (FSP) algorithms
\cite{burrage-etal06,munsky-khammash06-fsp,munsky-khammash07}
consider finite parts of the state space that can be reached either during the
whole time period of interest or during multiple time intervals into which the
time period of interest is split.
Then the computation of transient probabilities is conducted based on the
representation of the transient probability distribution as the product of
the initial probability distribution times a matrix exponential involving the
generator matrix of the underlying CTMC restricted to the finite projection.
However, computing matrix exponentials is well known to be an intricate issue
in itself \cite{moler-vanloan78,moler-vanloan03} and with FSP algorithms it
must be repeated multiple times, corresponding to repeated expansions
of the finite state projection.
Alternative state space truncation methods are based on adaptive wavelet
methods \cite{jahnke10,jahnke-udrescu10}, or on a conversion to discrete
time where it is dynamically decided which states to consider at a certain
time step in a uniformized discrete-time Markov chain \cite{andreychenko-etal18,mateescu-etal10}.

While examples show that these methods can handle the state space explosion in
some cases, for many systems they are still not feasible, because even the
considered finite part of the state space is too large, or because the specific
system dynamics hamper the computational methods involved.
In particular,
even if the state space is relatively small or can be reduced to a manageable
size, the stiffness problem must be handled satisfactorily and it is not clear
whether these methods are suitable for stiff systems.

In this paper, we present a novel numerical approximation method that tackles
both the state space explosion problem and the stiffness problem
where we consider efficient approximate numerical integration 
based on the fact that the CME can be cast in the form of a system of ordinary
differential equations (ODEs) and the computation of transient probabilities,
given an initial probability distribution, is an initial value problem (IVP).
We combine a dynamical state space truncation procedure, efficient numerical
integration schemes, and an adaptive step size selection based on local error
estimates. The dynamical state space truncation keeps the number of considered
states manageable while incurring only a small approximation error. It is much
more flexible and can reduce the state space much more than the aforementioned
methods. The use of efficient ODE solvers with adaptive step size control
ensures that the method is fast and numerically stable by taking as large as
possible steps without degrading the method's convergence order.

Our method approximates the solution of the CME by truncating large, possibly
infinite state spaces dynamically in an iterative fashion.
At a particular time instant $t$, we consider an approximation of the transient
probability distribution and
temporarily neglect states with a probability smaller than a threshold $\delta$,
that is, their probability at time $t$ is set to zero.
The CME is then solved for an (adaptively chosen) time step $h$ during which
the truncated state space is adapted to the probability distribution at time $t+h$.
More precisely, certain states that do not belong to the truncated part of
the state space at time $t$ are added at time $t+h$, when in the meantime they receive
a significant amount of probability which exceeds $\delta$. Other
states whose probabilities drop below $\delta$ are temporarily neglected.
The smaller the significance threshold $\delta$ is chosen the more accurate
the approximation becomes.

The dynamical state space truncation approach combined with a 4th-order 4-stage
explicit Runge-Kutta integration scheme was applied in
\cite{mikeev-etal11,mikeev-etal13} to the computation of certain transient rare
event probabilities in nonstiff models. The step size at each step was chosen
heuristically according to the smallest expected sojourn time in any
of the significant states, without any local error control. This works well for
nonstiff models, but stiff models require more sophisticated adaptive step size
selection strategies based on local error estimates and often even implicit
integration schemes \cite{butcher,hairer-wanner1,hairer-wanner2,shampine-etal-book03}.

In the present paper we substantially generalize and improve our previous work.
We extend the dynamical state space truncation to work for the whole class of
Runge-Kutta methods, explicit and implicit, each of which can be equipped with
an adaptive step size selection based on local error estimates.
This yields an accurate, numerically stable and computationally efficient
framework for 
the approximate solution of the CME for stochastic reaction networks
with extremely huge, possibly infinite, multidimensional state spaces.

In the next section we introduce the notation for stochastic reaction
networks, the underlying CTMC and in particular the CME that describes
the transient probability distribution.
Section~\ref{sec:numint} describes the dynamical state space truncation,
its efficient implementation for general Runge-Kutta methods, and the
adaptive step size selection based on local error estimates.
Examples are provided in Section~\ref{sec:examples}.
Finally, Section~\ref{sec:conclusion} concludes the paper and outlines
further research directions.

\section{Chemical Master Equation}
\label{sec:cme}
Consider a well-stirred mixture of $d\in{\mathbb N}$ molecular species
$S_1,\ldots,S_d$ in a thermally equilibrated system of fixed volume,
interacting through $M\in{\mathbb N}$ different types of chemical reactions,
also referred to as chemical reaction channels, $R_1,\ldots,R_M$.
At any time $t\geq 0$ a discrete random variable $X_k(t)$ describes the number
of molecules of species $S_k$ and the system state is given by the random
vector $X(t)=(X_1(t),\ldots,X_d(t))$.

The system changes its state due to one of the possible reactions, where
each reaction channel $R_m,m=1,\ldots,M$, is defined by a 
stoichiometric equation
\begin{equation}
\label{chem-reaction}
R_m:\
s_{m_1} S_{m_1} + \cdots + s_{m_r} S_{m_r} \stackrel{c_m}{\longrightarrow}
s_{m_{r+1}} S_{m_{r+1}} + \cdots + s_{m_\ell} S_{m_\ell} 
\end{equation}
with an associated stochastic reaction rate constant $c_m,$
reactants $S_{m_1},\ldots,S_{m_r},$ products $S_{m_{r+1}},\ldots,S_{m_\ell},$
and the corresponding stoichiometric coefficients $s_{m_1},\ldots,s_{m_\ell}\in\mathbb N,$
where $m_1,\ldots,m_\ell$ indexes those species that are involved in the reaction.
Mathematically, the stoichiometry is described by the state change vector
$v_m=(v_{m1},\ldots,v_{md}),$ where $v_{mk}$ is the change of molecules
of species $S_k$ due to $R_m.$
That is, if a reaction of type $R_m$ occurs when the system is in state $x,$
then the next state is $x+v_m,$ or, equivalently, state $x$ is reached, if a
reaction of type $R_m$ occurs when the system is in state $x-v_m.$

For each
reaction channel $R_m$ the reaction rate is given by a state-dependent
propensity function $\alpha_m,$ where $\alpha_m(x)dt$ is the conditional
probability that a reaction of type $R_m$ occurs in the time interval
$[t,t+dt),$ given that the system is in state $x$ at time $t.$ That is
\begin{equation}
\alpha_m(x)dt=P\left(\mbox{$R_m$ occurs in}\ [t,t+dt)\ |\ X(t)=x\right).
\end{equation}
The propensity function is given by $c_m$ times the
number of possible combinations of the required reactants and thus computes as
\begin{equation}
\label{def-propensity}
\alpha_m(x)=c_m \prod_{j=1}^{m_r} {x_{m_j} \choose s_{m_j}},
\end{equation}
where $x_{m_j}$ is the number of molecules of species $S_{m_j}$ present
in state $x$, and $s_{m_j}$ is the stoichiometric coefficient of $S_{m_j}$
according to (\ref{chem-reaction}). Because at any time the system's future
evolution only depends on the current state, $(X(t))_{t\geq 0}$ is a
time-homogeneous continuous-time Markov chain (CTMC), also referred to
as a Markov jump process, with
$d$-dimensional state space ${\cal X}\subseteq{\mathbb N}^d$.
Since the state space is countable it is always possible
to map it to $\N$, which yields a numbering of the states.

The conditional transient (time-dependent)
probability that the system is in state $x\in{\cal X}$ at time $t,$ given that
the system starts in an initial state $x_0\in{\cal X}$ at time $t_0,$ is
denoted by
\begin{equation}
\label{transprob}
p_x(t):=p(x,t):=p(x,t|x_0,t_0)=P\left(X(t)=x\ |\ X(t_0)=x_0\right)
\end{equation}
and the transient probability distribution at time $t$ is the collection of all
transient state probabilities at that time, represented by the row vector
$p(t)$.
The system dynamics in terms of the state probabilities' time
derivatives are described by the chemical master equation (CME)
\begin{equation}
\label{eq:master}
\frac{\partial p(x,t)}{\partial t}
= \sum_{m=1}^M \left(\alpha_m(x-v_m)p(x-v_m,t)
 -\alpha_m(x)p(x,t)\right)
 =:\mathcal Ap(t)(x),
\end{equation}
which is also well known as the system of Kolmogorov forward differential
equations for Markov processes \cite{vankampen}.
These stochastic reaction kinetics are physically well justified
since they are evidently in accordance with the theory of thermodynamics
\cite{gillespie92,vankampen}. In the thermodynamic limit the stochastic
description converges to classical deterministic mass action kinetics
\cite{kurtz72}.

Note that \eqref{eq:master} is the most common way to write the CME,
namely as a partial differential equation (PDE), where $t$ as well as
$x_1,\ldots,x_d$ are variables.
However, for any fixed state $x=(x_1,\ldots,x_d)$ the only free parameter
is the time parameter $t$ such that \eqref{eq:master} with fixed $x$
is an ordinary differential equation (ODE) with variable $t$.
In particular, when solving for the transient state probabilities numerical
ODE solvers can be applied.
We shall therefore use the notation $p_x(t)$ in the following.

\section{Approximate Numerical Integration of the CME}
\label{sec:numint}

Numerical integration methods for solving the CME,
given $p^{(0)}:=p(t_0)$, discretize the integration interval $[0,T]$ and
successively compute approximations
$p^{(1)}\approx p(t_1),p^{(2)}\approx p(t_2),\ldots,p^{(\eta)}\approx p(t_\eta)$,
where $0=t_0<t_1<t_2\cdots<t_\eta=T$ are the mesh points
and $h_i=t_{i+1}-t_i$ is the step size at the $i$-th step, $i=0,\ldots,{\eta-1}$.
With single-step methods each approximation $p^{(i+1)}\approx p(t_{i+1})$
is computed in terms of the previous approximation $p^{(i)}$ only, that is,
without using approximations $p^{(j)},j<i$.
For advanced methods the step sizes $h_i$ (and thus $\eta$, the number of steps) are not
determined in advance, but variable step sizes are determined in the course of
the iteration.

The system of ODEs described by the CME~\eqref{eq:master} is typically large or
even infinite, because there is one ODE for each state in the underlying CTMC,
that is, the size of the system of ODEs equals the size of the CTMC's state
space. Thus, its solution with standard numerical integration methods
becomes computationally infeasible.
However, one can exploit that at any time only a tractable number of
states have ``significant'' probability, that is, only relatively few
states have a probability that is greater than a small threshold.

The main idea of our dynamical state space truncation for numerical
integration methods is to integrate only those differential equations in the
CME~\eqref{eq:master} that correspond to significant states.
All other state probabilities are (temporarily) set to zero.
This reduces the computational effort significantly since in each iteration
step only a comparatively small subset of states is considered.
Based on the fixed probability threshold $\delta>0$, we dynamically decide
which states to drop or add, respectively.
Due to the regular structure of the CTMC the approximation error of the
algorithm remains small since probability mass is usually concentrated
at certain parts of the state space. The farther away a state is from
a ``significant set'' the smaller is its probability.
Thus,
in most cases
the total error of the approximation remains small.
Since in each iteration step probability mass may be
 ``lost'' the approximation error at step $i$ is
the sum of all probability mass lost (provided that the numerical
integration could be performed without any  errors), that is,
\begin{equation}
1-\sum_{x\in{\cal S}}p_x^{(i)}.
\end{equation}

It is important to note that other than static state space truncation
approaches our dynamical approach allows that in the course of the
computation states can ``come and go", that is, states join the significant
set if and only if their current probability is above the threshold
$\delta$ and states in the significant set are dropped immediately when
their current probability falls below $\delta$.
Furthermore, states that have previously been dropped may come back, that is
they are re-considered as significant as soon as they receive a probability
that exceeds the threshold $\delta$.
This is substantially different from state space exploration techniques where
only the most probable states are generated but states are never dropped as
time progresses like for instance in \cite{desouza-ochoa92} with regard to
approximating stationary distributions.
Our dynamical state space truncation approach is also
much more flexible than finite state projection (FSP) algorithms
\cite{burrage-etal06,munsky-khammash06-fsp,munsky-khammash07}
which work over pre-defined time intervals with the same subset of states,
where in particular for stiff systems many reactions can occur during any
time interval, so that in order to safely meet reasonable accuracy requirements
the resulting subset of states is often still extremely large.
In contrast to that we update our set of significant states in each adaptively chosen
time step,
without much overhead.
Furthermore, by numerically integrating the ODEs we avoid the
intricate computation of matrix exponentials required in FSP algorithms and
by using an efficient data structure we do not even need to generate any
matrices.

In order to avoid the explicit construction of a matrix
and in order to work with a dynamic set $\sig$ of significant states
that changes in each step, we use for a state $x$ a data structure
 with the following components:
\begin{itemize}
\item fields $x.\prob$, $x.\nextprob$ for the approximated probabilities
      $p^{(i)}_x$ and $p^{(i+1)}_x$, respectively,
\item for all $m$ with $\alpha_m(x)>0$ 
 a pointer to the successor state $x+v_m$
as well as a field with the rate $\alpha_m(x)$.
\end{itemize}

\begin{algorithm}[t]
\caption{A numerical adaptive step size integration scheme.}
\label{alg:fApprox}
\begin{algorithmic}[1]
\SLet{$\sig$}{$\{x: p_x(0)>\delta\}$}
\State \Let{$t$}{0}, \Let{$i$}{0}
\State compute $h_0$
\While {$t<T$}
\State compute $p^{(i+1)}$
\State compute $\hat{\epsilon}_i$, $h_{i+1}$
\If {step successful}
\State update $\sig$
\State \Let{$t$}{$t+h_i$}, \Let{i}{i+1}
\EndIf
\EndWhile
\end{algorithmic}
\end{algorithm}

The workflow of the numerical integration scheme is given in pseudocode
in Algorithm~\ref{alg:fApprox}. We start at time $t=0$ and initialize the set $\sig$
as the set of all states that have initially a probability greater than $\delta$.
We compute the initial time step $h_0$. In each iteration step we compute the
approximation $p^{(i+1)}$ using an explicit or implicit Runge-Kutta method
(see Sections~\ref{sec:RK-explicit}, \ref{sec:RK-implicit}).
We check whether the iteration step was successful computing the
local error estimate $\hat{\epsilon_i}$ and ensuring that error tolerance
conditions are met. If so, then for each state we update the field $x.\prob$
with $x.\nextprob$, and remove the state from $\sig$ if its probability becomes less than $\delta$.
Based on the local error estimate, we choose a time step for the next iteration
(or the repetition of the iteration in case it was not successful). This and the computation
of the initial time step is detailed in Section~\ref{sec:stepsize}.

\subsection{Runge-Kutta methods}
\label{sec:RK-explicit}

We consider the whole family of Runge-Kutta methods, which
proceed each time step of given step size in multiple {\em stages}.
More precisely, a general $s$-stage Runge-Kutta method proceeds according to
the iteration scheme
\begin{equation}
p^{(i+1)}=p^{(i)}+h_i\sum_{\ell=1}^s b_\ell k^{(\ell)},
\label{eq:fRK}
\end{equation}
\begin{equation}
k^{(\ell)}:={\cal A}\left(p^{(i)}+h_i\sum_{j=1}^s a_{\ell j}k^{(j)}\right),
\label{eq:fRK-stages}
\end{equation}
which is uniquely defined by weights
$b_1\ldots,b_s>0$ with $b_1+\cdots+b_s=1$ and coefficients
$0\leq a_{\ell j}\leq 1$, $\ell=1,\ldots,s,\ j=1,\ldots,s$.
Thus, $k^{(\ell)}$ is an approximation to $p(t_i+h_i c_\ell)$,
where $c_\ell=a_{\ell 1}+\cdots+a_{\ell s}$, and $k^{(\ell)}(x)$ is the
component of the vector $k^{(\ell)}$ that corresponds to state $x$.
Hence, $k^{(\ell)}(x)$ is the probability of $x$ at stage $\ell$.
If $a_{\ell j}=0$ for all $j\geq\ell$, then the sum in Equation~\eqref{eq:fRK-stages}
effectively runs only from $1$ to $\ell-1$, which means that for each
$\ell=1,\ldots,s$ the computation of $k^{(\ell)}$ includes only previous stage
terms $k^{(j)},j<\ell$. Therefore, $k^{(1)},\ldots,k^{(s)}$ can be computed
sequentially, that is, $a_{\ell j}=0$ for all $j\geq\ell$ yields explicit
integration schemes. If there is at least one $j\geq\ell$ with $a_{\ell j}>0$,
then the integration scheme is implicit, which implies that the solution of at
least one linear system of equations is required per iteration step.

\begin{algorithm}[t]
\caption{A single iteration step of a general explicit $s$-stage Runge-Kutta scheme,
defined by $s,b_1,\ldots,b_s$, and $a_{\ell j}$ for $\ell=1,\ldots,s,\ j<\ell$,
with dynamical state space truncation, which approximates the solution of
the CME.}
\label{alg:fRK-explicit}
\begin{algorithmic}[1]
\For{\Let{$\ell$}{1} to $s$}
\ForAll{$x \in \sig$}
\State \Let{$\hat{p}$}{$x.\prob + h\cdot\sum_{j=1}^{\ell-1} a_{\ell j} \cdot x.k_j$}
\ForAll{$m \in \{1,\ldots,M\}$ : $x\!+\!v_m\geq\zvec ~\textbf{and}~$\par
\hskip\algorithmicindent\hskip\algorithmicindent\hskip\algorithmicindent $\left(x+v_m \in \sig ~\textbf{or}~ h\cdot\alpha_m(x)\cdot\hat{p}>\tilde{\delta} \right)$}
\IIf {$x+v_m \not\in \sig$} \Let{$\sig$}{$\sig\cup\{x+v_m\}$} \EndIIf
\State \Let{$x.k_\ell$}{$x.k_\ell - h\cdot\alpha_m(x)\cdot\hat{p}$}
\State \Let{$(x+v_m).k_\ell$}{$(x+v_m).k_\ell + h\cdot\alpha_m(x)\cdot\hat{p}$}
\EndFor{}
\EndFor{}
\EndFor{}
\ForAll{$x \in \sig$}
\State \Let{$x.\nextprob$}{$x.\prob + h\cdot\sum_{j=1}^s b_j \cdot x.k_j$}
\State \Let{$x.k_1$}{0}, \ldots, \Let{$x.k_s$}{0}
\EndFor{}
\end{algorithmic}
\end{algorithm}

A single iteration step for general explicit $s$-stage Runge-Kutta schemes
is given in pseudocode in Algorithm~\ref{alg:fRK-explicit}. Note that for each state
besides $x.\prob$, $x.\nextprob$, we additionally store fields $x.k_1,\ldots, x.k_s$
for the stage terms $k^{(1)}(x), \ldots, k^{(s)}(x)$ and initialize them with zero.
We compute the approximation of $p^{(i+1)}$ based on Equation~\eqref{eq:fRK}
by traversing the set $\sig$ $s+1$ times. In the first $s$ rounds (lines 1-10)
we compute $x.k_1,\ldots,x.k_s$ according to Equation~\eqref{eq:fRK-stages} and
in the final round (lines 11-14) we accumulate the summands and zero $x.k_1,\ldots,x.k_s$.
While processing state $x$ in round $\ell\leq s$, for each reaction channel $m$, we transfer probability mass from state $x$ to its successor $x+v_m$,
by subtracting a term from $x.k_i$ and adding the same term to $(x+v_m).k_i$ (lines 6-7).
We do so after checking (line 4) whether $x+v_m$ is already in $\sig$, and if not, whether
it is worthwhile to add $x+v_m$ to
$\sig$, that is, we guarantee that $x+v_m$ will receive enough probability
mass and that $x+v_m$ will not be removed in the same iteration.
Thus, we add $x+v_m$ to $\sig$ (line 5) only if the inflow
$h \cdot \alpha_m(x) \cdot (x.\prob + h\cdot\sum_{j=1}^{\ell-1} a_{\ell j} \cdot x.k_j)$
to $x+v_m$ is greater or equal than a certain threshold $\tilde\delta>0$.
Obviously,  $x+v_m$ may receive more probability mass from other states
and the total inflow may be greater than $\tilde\delta$. Thus, if a state is
not a member of $\sig$ and if for each incoming transition the inflow
probability is less than $\tilde\delta$, then this state will not be added to
$\sig$ even if the total inflow is greater or equal than $\tilde\delta$. This
small modification yields a significant speed-up since otherwise all states that
are reachable within at most $s$ transitions will always be added to $\sig$,
but many of the newly added states will be removed in the
same iteration.

\subsection{Implicit Methods}
\label{sec:RK-implicit}

The advantage of implicit methods is that they can usually take larger (and thus
fewer) steps, which comes at the price of an increased computational effort
per step, but paying that price can lead to large speed-up of the overall
integration over the time interval $[0,T]$.
It is common sense that in general the efficient solution of stiff ordinary
differential equations requires implicit integration schemes
\cite{hairer-wanner2}.
In the present paper, with regard to implicit numerical integration we restrict
ourselves to the implicit Euler method, hence the special case of a
one-stage Runge-Kutta method with $a_{11}=1$ and $b_1=1$, which yields
\begin{equation}
p^{(i+1)}=p^{(i)}+h_i{\cal A}p^{(i+1)}
\end{equation}
and requires to solve the linear system
\begin{equation}
p^{(i+1)}-h_i {\cal A}p^{(i+1)} = p^{(i)}
\label{eq:linsys-euler}
\end{equation}
for $p^{(i+1)}$ in each step.

Of course, when considering a standard approach to the numerical integration of
the CME, where no state space truncation is considered, then this linear system
is huge, possibly infinite, and its solution is often impossible. In conjunction
with our dynamical state truncation procedure, the linear system is reduced
similarly to the reduction of the state space and the number of differential
equations to be integrated per step, respectively. However, there a subtleties
that must be properly taken into account.

Firstly, since we do not need to maintain
huge matrices but we use the previously described dynamical data structure
the solution of the linear system must be accordingly implemented with this
data structure. Secondly, some states that are not significant at time $t$ may
receive a significant probability at time $t+h_i$ and must be included in the
linear system. Thus, a dynamical implementation of the solution of the linear
system is required.
Therefore, iterative solution techniques for linear systems that are usually
simply 
defined by a fixed matrix must be properly
adapted to the dynamical data structure and the dynamical state space truncation.

In fact, when using implicit numerical integration schemes in conjunction with
the dynamical state space truncation procedure, in principle the solution of a
linear system in each integration step is a challenging potential bottleneck.
It is therefore a key point and a key contribution to implement it efficiently.

\begin{algorithm}[t]
\caption{A single iteration step of an implicit Euler scheme using the Jacobi method with dynamical state space truncation, which approximates the solution of the CME.}
\label{alg:fBEuler}
\begin{algorithmic}[1]
\While {convergence not reached}
\ForAll{$x \in \sig$}
\ForAll{$m \in \{1,\ldots,M\}$ : $x\!+\!v_m\geq\zvec ~\textbf{and}~$\par
\hskip\algorithmicindent\hskip\algorithmicindent\hskip\algorithmicindent $\left(x+v_m \in \sig ~\textbf{or}~ h\cdot\alpha_m(x)\cdot x.p_1>\tilde{\delta} \right)$}
\IIf {$x+v_m \not\in \sig$} \Let{$\sig$}{$\sig\cup\{x+v_m\}$} \EndIIf
\State \Let{$(x+v_m).k$}{$(x+v_m).k + \alpha_m(x)\cdot x.p_1$}
\EndFor{}
\EndFor{}
\ForAll{$x \in \sig$}
\State \Let{$x.p_2$}{$(x.p + h \cdot x.k) / (1 + h \cdot \alpha_0(x))$}
\State check convergence for state $x$
\State \Let{$x.p_1$}{$x.p_2$}
\State \Let{$x.k$}{0}
\EndFor{}
\EndWhile
\end{algorithmic}
\end{algorithm}

We illustrate the solution of the linear system \eqref{eq:linsys-euler} using the Jacobi method, which yields the following iterative scheme
\begin{equation}
\label{eq:beuler_jacobi}
p^{(i+1,j+1)}_x = \frac{p^{(i)}_x+h_i \cdot \sum_{m=1}^{M} \alpha_m(x-v_m) \cdot p^{(i+1,j)}_{x-v_m}} {1 + h_i \cdot \alpha_0(x)},
\end{equation}
where $\alpha_0(x)=\sum_{m=1}^{M} \alpha_m(x)$. 
The pseudocode is given in Algorithm~\ref{alg:fBEuler}.
In the $(i+1)$-th iteration of the adaptive numerical integration scheme (Algorithm~\ref{alg:fApprox}),
we store the ``old'' approximation of the state probability $p^{(i+1,j)}_x$ in the field $x.p_1$ and the ``new'' approximation $p^{(i+1,j+1)}_x$ in the field $x.p_2$.
We initialize $x.p_1$ with the state probability from the $i$-th iteration $p^{(i)}_x$.
In lines 2-7 for each state we compute the sum $\sum_{m=1}^{M} \alpha_m(x-v_m) \cdot p^{(i+1,j)}_{x-v_m}$
and store it in a field $x.k$. While processing state $x$, for each reaction channel $m$, we transfer probability mass from state $x$ to its successor $x+v_m$. Similarly to Algorithm~\ref{alg:fRK-explicit}, we only add a new
state to $\sig$ if it receives enough probability mass.
In lines 9-14 for each state we compute the ``new'' approximation of $p^{(i+1)}_x$ according to \eqref{eq:beuler_jacobi}
and check whether the convergence criterion
\begin{equation}
\label{eq:conv_crit}
|p^{(i+1,j+1)}_x-p^{(i+1,j)}_x| \leq \max(\lsrtol\cdot\max(p^{(i+1,j+1)}_x,p^{(i+1,j)}_x),\lsatol),
\end{equation}
is fulfilled
for some relative and absolute tolerances $\lsrtol>0$ and $\lsatol>0$. Algorithm~\ref{alg:fBEuler}
terminates if \eqref{eq:conv_crit} holds for all states $x\in\sig$. After the convergence of the Jacobi method,
the field $x.\nextprob$ contains the approximation $p^{(i+1)}_x$.

For our numerical experiments we use the Gauss-Seidel method, which is known to converge faster
than the Jacobi method. The iterative solution is given as
\begin{equation}
\label{eq:beuler_gauss_seidel}
p^{(i+1,j+1)}_x = \frac{p^{(i)}_x+h_i \cdot \sum_{m=1}^{M} \alpha_m(x-v_m) \cdot
\left[ \xi_{x-v_m}\cdot p^{(i+1,j+1)}_{x-v_m} + (1-\xi_{x-v_m})\cdot p^{(i+1,j)}_{x-v_m} \right] } {1 + h_i \cdot \alpha_0(x)},
\end{equation}
where $\xi_{x-v_m}$ is an indicator (or flag) that takes the value $1$ if the
state $x-v_m$ has been already processed, and $0$ otherwise.
Thus, in \eqref{eq:beuler_gauss_seidel} in the summation, we use the ``new''
approximations of the processed states and the ``old'' probability if the
approximation was not yet updated in the current iteration.
We modify Algorithm~\ref{alg:fBEuler} as follows.
We compute the sum as before, but after processing a state $x$ in line 9,
we mark it as processed and propagate $\alpha_m(x)\cdot\left(x.p_2-x.p_1\right)$
to the successor states which are not marked as processed. After this,
the field $x.k$ contains the sum required for \eqref{eq:beuler_gauss_seidel}.

\subsection{Local Error Control and Adaptive Step Size Selection}
\label{sec:stepsize}

The accuracy as well as the computing time of numerical integration methods
depend on the order $p$ of the method and the step size. The error in a single
step with step size $h$ is approximately $ch^{p+1}$ with a factor $c$ that
varies over the integration interval.
Hence, one crucial point for the efficiency of numerical integration methods
is the step size selection. It is well known that methods with constant
step size perform poorly if the solution varies rapidly in some parts of the
integration interval and slowly in other parts
\cite{butcher,hairer-wanner1,hairer-wanner2,shampine-etal-book03}.
Therefore, adaptive step size selection so that the accuracy and the
computing time are well balanced is highly desirable for explicit and
implicit integration schemes. For both classes of schemes we base
our step size selection strategy on local error estimates.

Our goal is to control the local error and, accordingly, to choose the step
size so that at each step $i$ for all states $x\in\sig$,
\begin{equation}
|p_x(t_i)-p_x^{(i)}|\leq\rtol\cdot|p_x(t_i)|+\atol,
\end{equation}
where $\rtol$ and $\atol$ are user-specified relative and absolute
error tolerances. In particular,
note that we use a {\em mixed error control}, that is, a criterion that
accounts for both the relative and the absolute error via corresponding
relative and absolute error tolerances,
because in practice using either a pure relative error control or a pure
absolute error control can cause serious problems, see, e.g.,
Section~1.4 of
\cite{shampine-etal-book03}.
Of course, the true local errors are not available and we must estimate
them along with each integration step.

For the explicit and the implicit Euler method
we compute a local error estimate similarly to the {\em step doubling}
approach, that is, we approximate $p^{(i+1)}$ by taking the time step
$h_i$ and independently taking two consecutive time steps of length
$h_i/2$.
The local error estimate is then the vector
\begin{equation}
\heps^{(i)} = p^{(i+1),(h_i)}-p^{(i+1),(h_i/2)}
\end{equation}
with components $\heps_x^{(i)}, x\in\sig$, where $p^{(i+1),(h_i)}$ and
$p^{(i+1),(h_i/2)}$ denote the approximations computed with time step $h_i$
and with two consecutive time steps of length $h_i/2$, respectively.

The {\em embedded} Runge-Kutta methods provide an alternative way for the step size
control. Along with the approximation of order $p$, they deliver the approximation of
order $p-1$ computed as
\begin{equation}
\tilde{p}^{(i+1)}=p^{(i)}+h_i\sum_{\ell=1}^s b^*_\ell k^{(\ell)},
\label{eq:fRK-embedded}
\end{equation}
where $b^*_1\ldots,b^*_s>0$ with $b^*_1+\cdots+b^*_s=1$.
Then the local error estimate is the vector
\begin{equation}
\heps^{(i)}=p^{(i+1)}-\tilde{p}^{(i+1)}\ =\ h_i\sum_{\ell=1}^s (b_\ell-b_\ell^*)k^{(\ell)}
\end{equation}
with components $\heps_x^{(i)}=p_x^{(i+1)}-\tilde{p}_x^{(i+1)}, x\in\sig$.

Now, with regard to the step size selection assume we have made a `trial' step
with a given step size $h_i$ and computed the corresponding local error estimate.
Then we accept the step if for all significant states the
local error estimate is smaller than the prescribed local error
tolerance. More precisely,
\begin{equation}
\forall x\in\sig: |\heps_x^{(i)}| \leq \max(\rtol\cdot\max(p_x^{(i)},p_x^{(i+1)}),\atol)=:\tau_x^{(i)},
\label{localerror-crit}
\end{equation}
which implies
\begin{equation}
\forall x\in\sig: |\heps_x^{(i)}| \leq\rtol\cdot\max(p_x^{(i)},p_x^{(i+1)})+\atol.
\end{equation}
If the step is not accepted, then we have to decrease the step size.
Otherwise we can proceed to the next step where it is likely that we can use
an increased step size, because the current one might be smaller
than necessary. In both cases, acceptance or rejection, we have to specify
by how much the step size is decreased or increased, respectively, and in
both cases we do this based on the local error estimate.

Define $\teps_x^{(i)}:=\heps_x^{(i)}/\tau_x^{(i)},x\in\sig$, denote
by $\teps^{(i)}$ the corresponding vector containing the components
$\teps_x^{(i)}$, and define
\begin{equation}
\alpha:=\sqrt[p+1]{\frac{1}{||\teps^{(i)}||_\infty}}.
\end{equation}
It can be easily seen that the largest step size that yields
a local error estimate satisfying \eqref{localerror-crit} can be
approximated by $h_i\alpha$.
Note that $\alpha<1$ if \eqref{localerror-crit} is satisfied and
$\alpha\geq 1$ otherwise. This means we can use $\alpha$ as a factor in
both possible cases, that is, for a too large step size that has been
rejected and must be decreased for a retrial, and for an accepted step size to
set an increased step size for the next step.
In practice we also have to account for the fact that the local error is only
estimated. Rejecting a step and re-computing it with a smaller step size
should be avoided as much as possible. Therefore, rather than $\alpha$
we consider $\rho\alpha$ with a safety factor $\rho<1$.
Besides, step sizes must not be too large and also too large changes of
the step size must be avoided since otherwise the above approximation of the largest
possible step size is not valid \cite{shampine-etal-book03}. 
If the step with step size $h_i$ is accepted, then we set
\begin{equation}
h_{i+1}:=\min(h_{\max},h_i\max(5,\rho\alpha))
\end{equation}
as the initial trial step size for the next step, where $h_{\max}=0.1\cdot T$. Otherwise, we decrease the
step size for the current step according to
\begin{equation}
h_i := \max(h_{\min},h_i\max(0.1,\rho\alpha)),
\end{equation}
where $h_{\min}=16\cdot\epsilon(t_i)$ and $\epsilon(t_i)$ is the absolute distance
between $t_i$ and the next floating-point number of the same precision as $t_i$.
So, $h_{\min}$ is such that $t_i$ and $t_i + h_{\min}$ are different in working precision.

For the computation of the initial trial step size, we first compute
$\mathcal Ap(t_0)$ (see~\eqref{eq:master}). This can be done using one stage of Algorithm~\ref{alg:fRK-explicit}.
Then we compute
\begin{equation}
h_0 = \max\left(h_{\min},\min\left(h_{\max}, \rho\cdot\frac{\sqrt[p+1]{\rtol}}{\rtol\cdot\underset{x\in\sig}{\max}{\displaystyle\frac{|x.k_1|}{\max\left(\rtol\cdot x.p,\atol\right)}}}\right)\right).
\end{equation}

In the $i$-th iteration, we stretch the time step $h_i$, if it lies within 10\% of $T-t_i$.
Thus, we set the time step $h_i$ to $\left(T-t_i\right)$ if $1.1\cdot h_i \geq T-t_i$. Note that
this also covers the case when the time step $h_i$ is too large, and using it would lead to
jumping over the final time point $T$.

\section{Numerical examples}
\label{sec:examples}

In this section, we present numerical examples in order to demonstrate the suitability
of our approach, its accuracy, run time and the number of significant states to
be processed corresponding to the number of differential equations to be integrated.
As our first example we consider a birth-death process for which analytical solutions
are available such that we can indeed compare our numerical results with exact values.
Then we consider a more complex yeast cell polarization model.

We compare the accuracies, run times and numbers of significant states of explicit
Euler (referred to as `\euler' in the following figures and tables), implicit (backward)
Euler (`\beuler'), and an embedded Runge-Kutta (`\rkfourfive') with weights and coefficients
chosen according to \cite{dormand-prince80}
together with local error control and adaptive step size selection as described in the
previous section. Note that \rkfourfive is similar to the {\tt ode45} method of MATLAB,
but while with MATLAB's {\tt ode45} only systems of ODEs of moderate size can be
solved, here, of course, we consider it in conjunction with our dynamical state space
truncation procedure.

For our numerical experiments we fix the relative tolerance $\rtol=10^{-3}$.
For the dynamical state space truncation, we use $\delta=\atol$, which agrees
with the error control property of the ODE solution that the components smaller than $\atol$
are unimportant.
As a safety factor in the time step selection procedure we use $\rho=0.8$.
In the solution of the linear system required for the implicit Euler method we set
$\lsrtol = \rtol$ and $\lsatol = \atol$.


\subsection{Birth-Death Process}

Our first example is the birth-death process given as
\[
\emptyset \xrightleftharpoons[c_2]{c_1} S_1,
\]
with $S_1$ as the only species and propensity functions
$\alpha_1(x) = c_1, \alpha_2(x) = c_2 x_1$.
It is clear that the state space is the infinite set ${\mathbb N}$ of all nonnegative integers
so that the corresponding system of differential equations is infinite, too.
We chose the rate constants $c_1=1$, $c_2=0.1$, the initial state $x_1(0)=1000$
and final time horizon $T=50$. We analyze the model with different values of
$\atol\in\{10^{-10},10^{-12},10^{-14}\}$.
Since reporting the probabilities of single states over time is not of any
practical interest we focus on representative properties and on informative
measures of the accuracy and the efficiency of our method.

\begin{figure}[h]
\centering
\subfloat[][]{\includegraphics[width=0.33\textwidth]{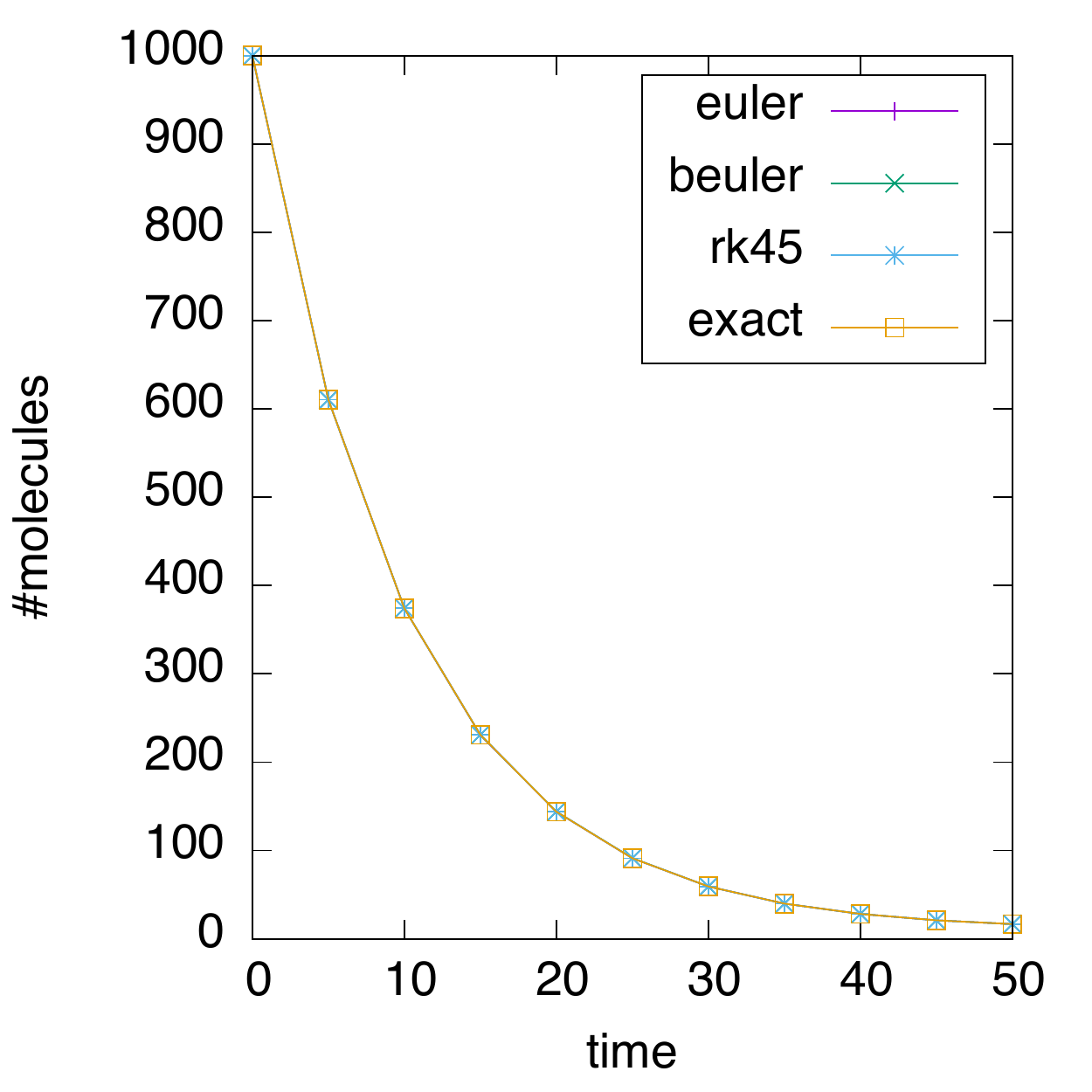}}
\subfloat[][]{\includegraphics[width=0.33\textwidth]{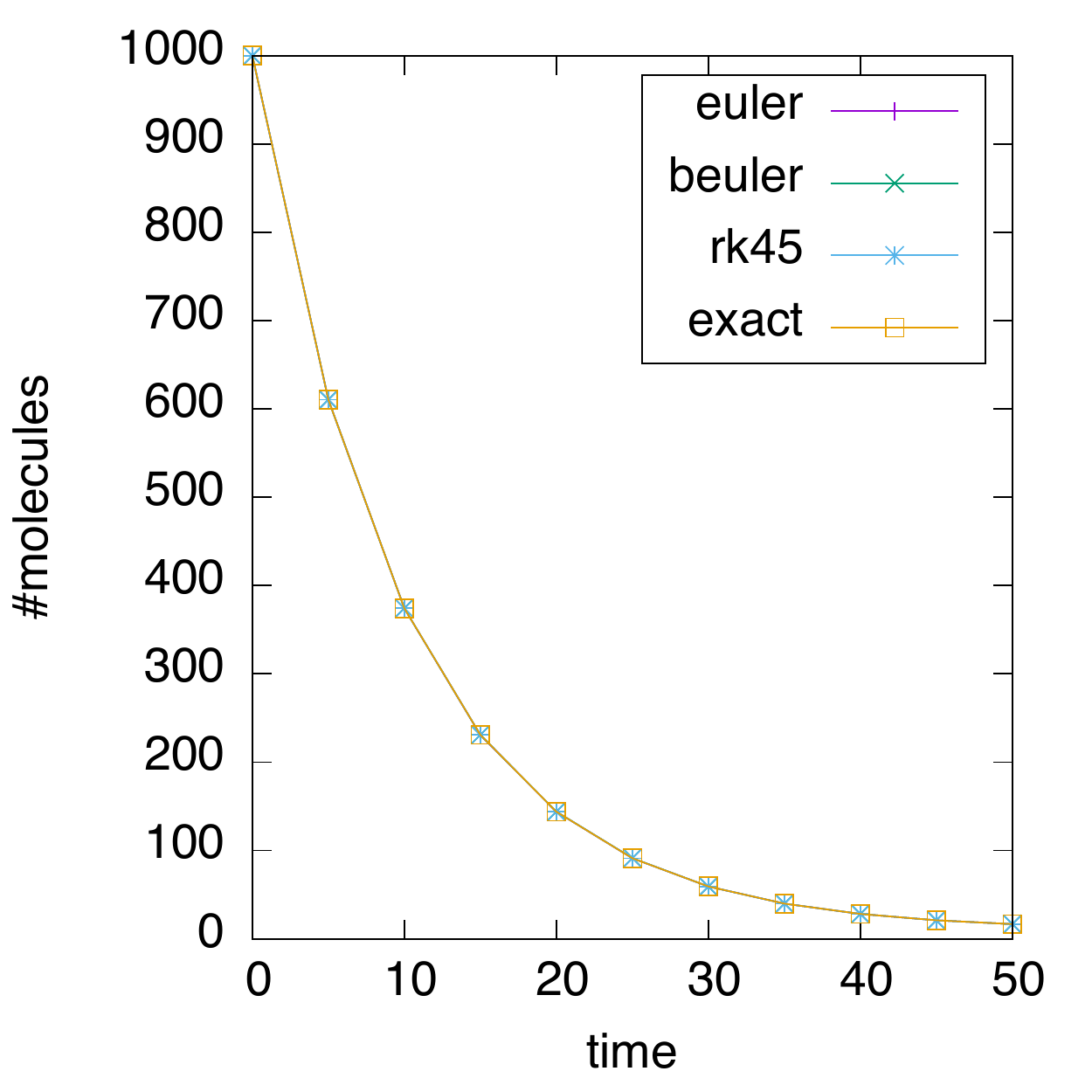}}
\subfloat[][]{\includegraphics[width=0.33\textwidth]{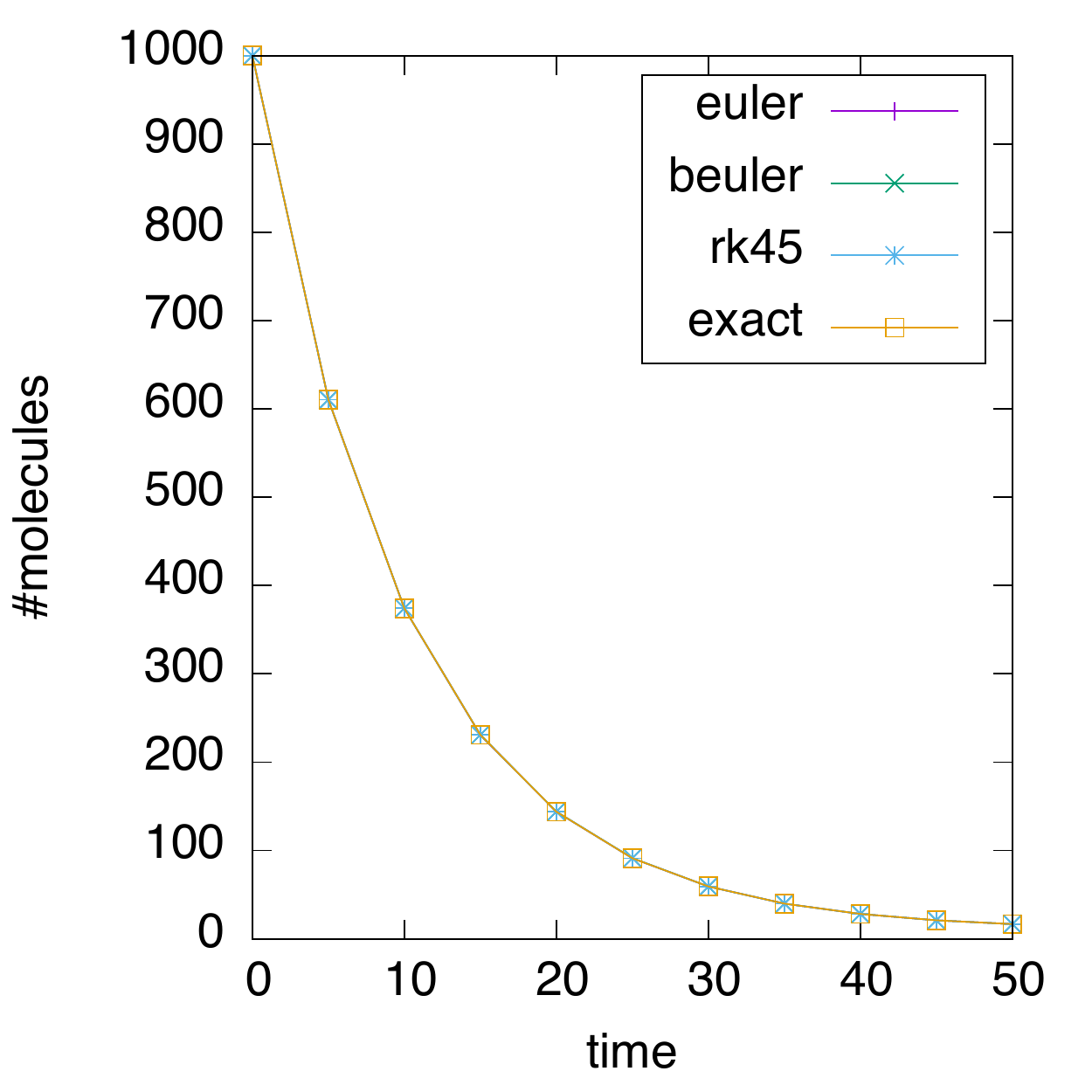}}
\caption{Average species count for birth-death process ($\atol=$ (a) $10^{-10}$, (b) $10^{-12}$, (c) $10^{-14}$). }
\label{fig:bd1_means}
\end{figure}

In Figure~\ref{fig:bd1_means} we plot the average number of species $S_1$ over
time as obtained with our approximate numerical integration schemes, where
the run times were less than one second.
We also plot
the exact solution obtained according to \cite{jahnke-huisinga07}. The plots
show for all considered values of $\atol$ that there is no visible difference
between the exact values and our approximations, which suggests that our
approximations are indeed extremely accurate.

\begin{figure}[h!]
\centering
\subfloat[][]{\includegraphics[width=0.33\textwidth]{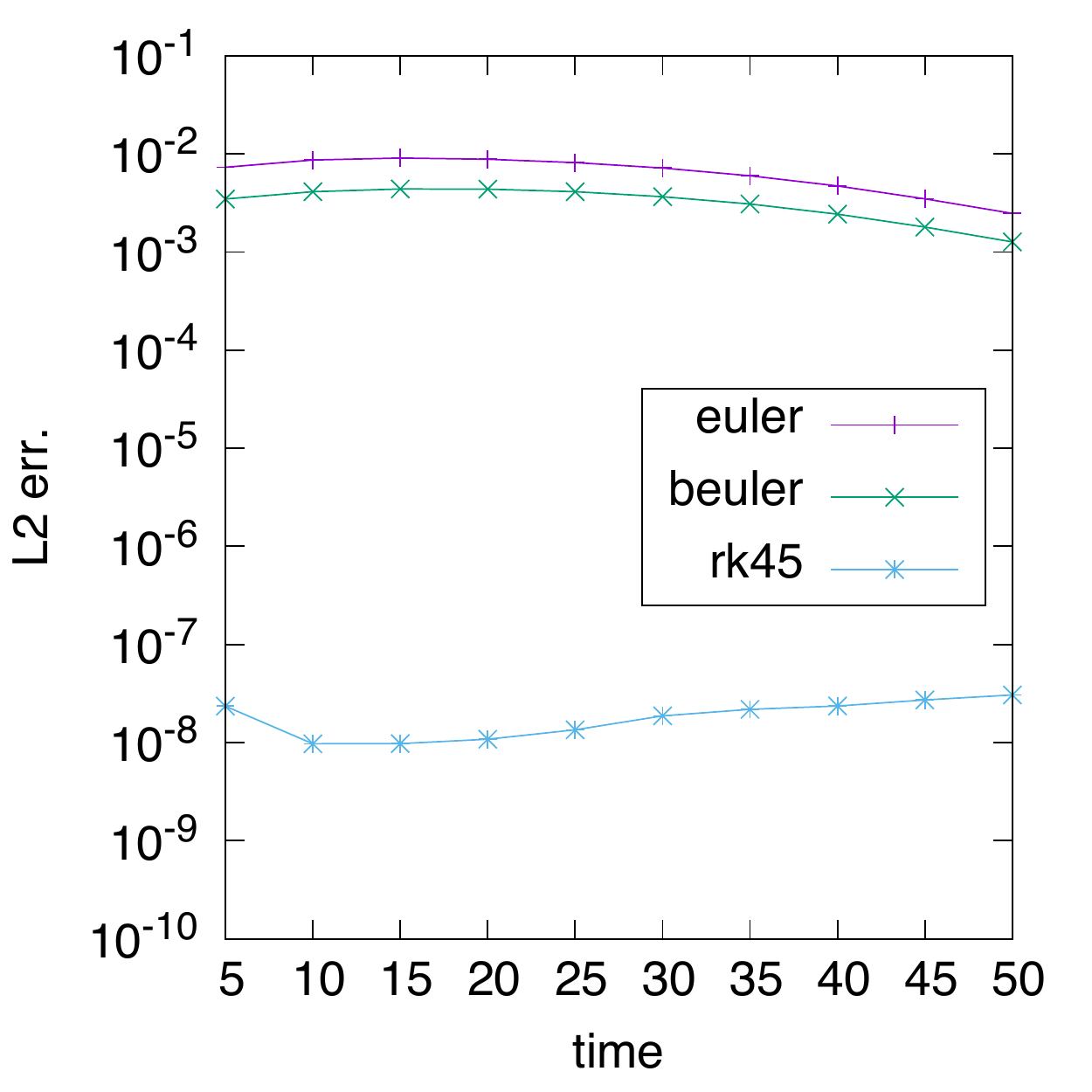}}
\subfloat[][]{\includegraphics[width=0.33\textwidth]{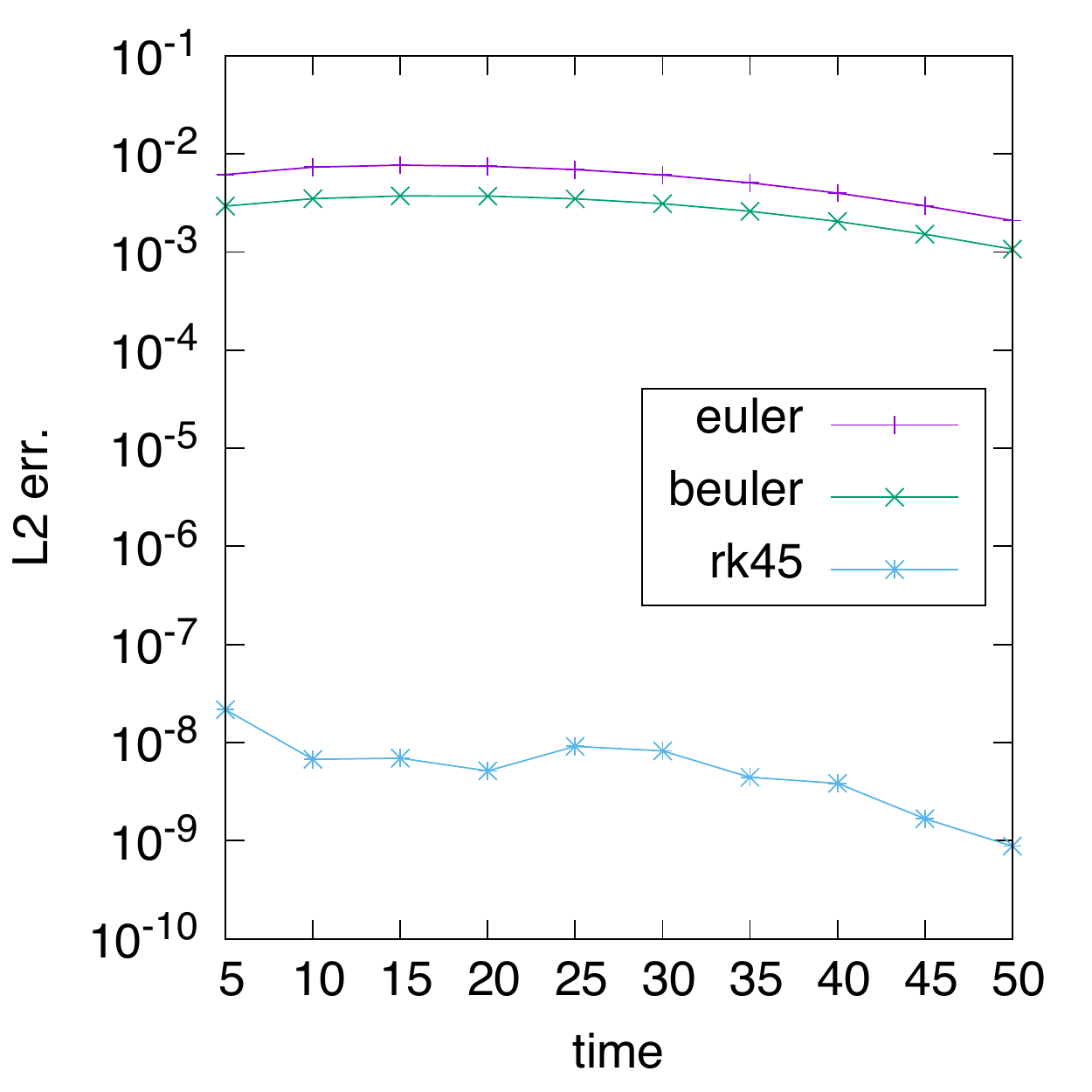}}
\subfloat[][]{\includegraphics[width=0.33\textwidth]{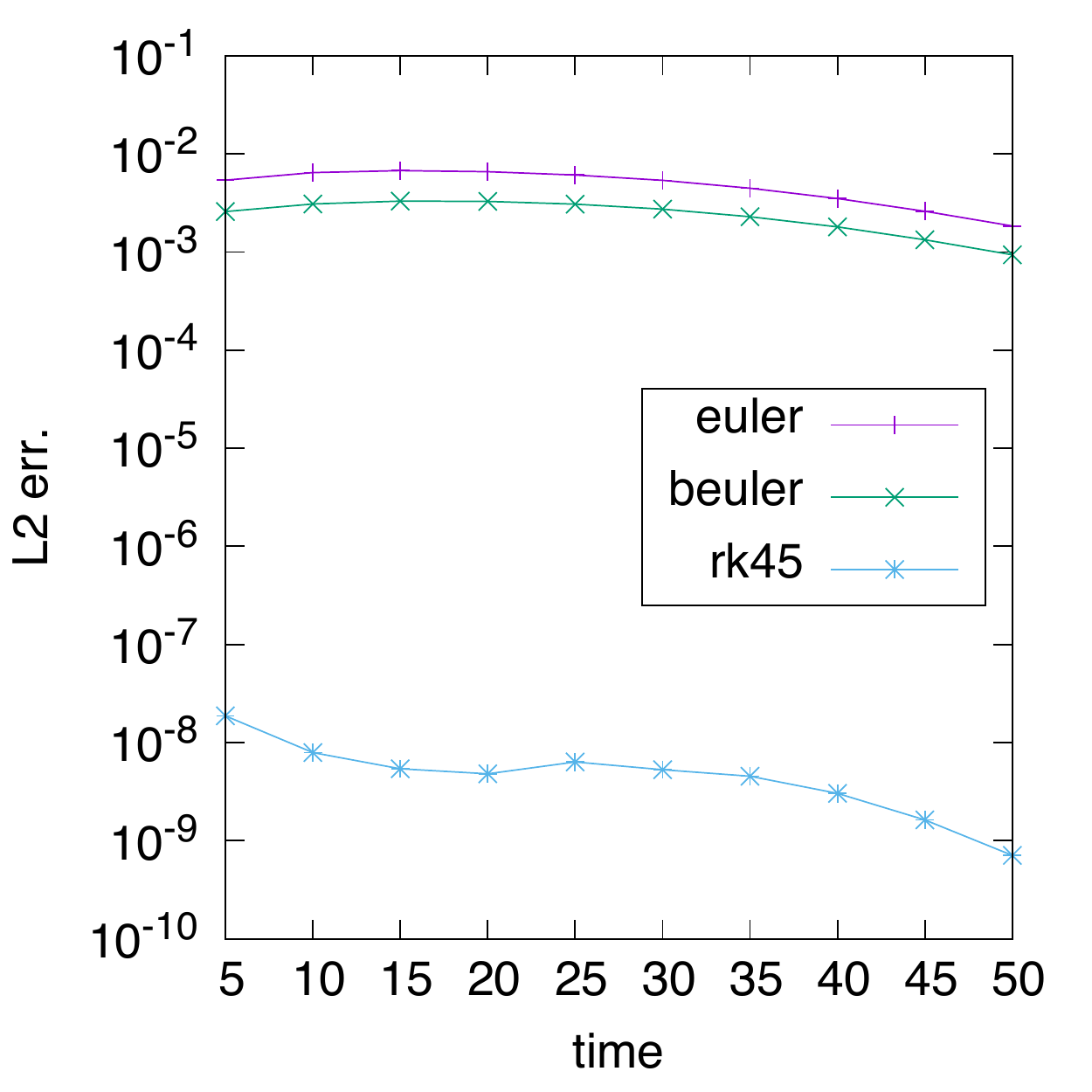}}
\caption{L2 error for birth-death model ($\atol=$ (a) $10^{-10}$, (b) $10^{-12}$, (c) $10^{-14}$).}
\label{fig:bd1_err_distr}
\end{figure}

Figure~\ref{fig:bd1_err_distr} depicts plots of the L2 error
$$
||p_x(t_{\eta})-p_x^{(\eta)}||_2
$$
with $t_{\eta}=T=50$.
The norm is computed over all states with positive probabilities in the exact solution.
Note that if there is no corresponding state in $\sig$, its probability is taken as 0.

It can be seen that in all cases the L2 error is less than $10^{-2}$, which confirms the
high accuracy of the approximations also formally. It can be further seen that for this
example \euler is slightly more accurate than \beuler and that \rkfourfive is even by
orders of magnitude more accurate than the Euler schemes.
This is well in accordance with the higher order of \rkfourfive.

\begin{figure}[h]
\centering
\subfloat[][]{\includegraphics[width=0.33\textwidth]{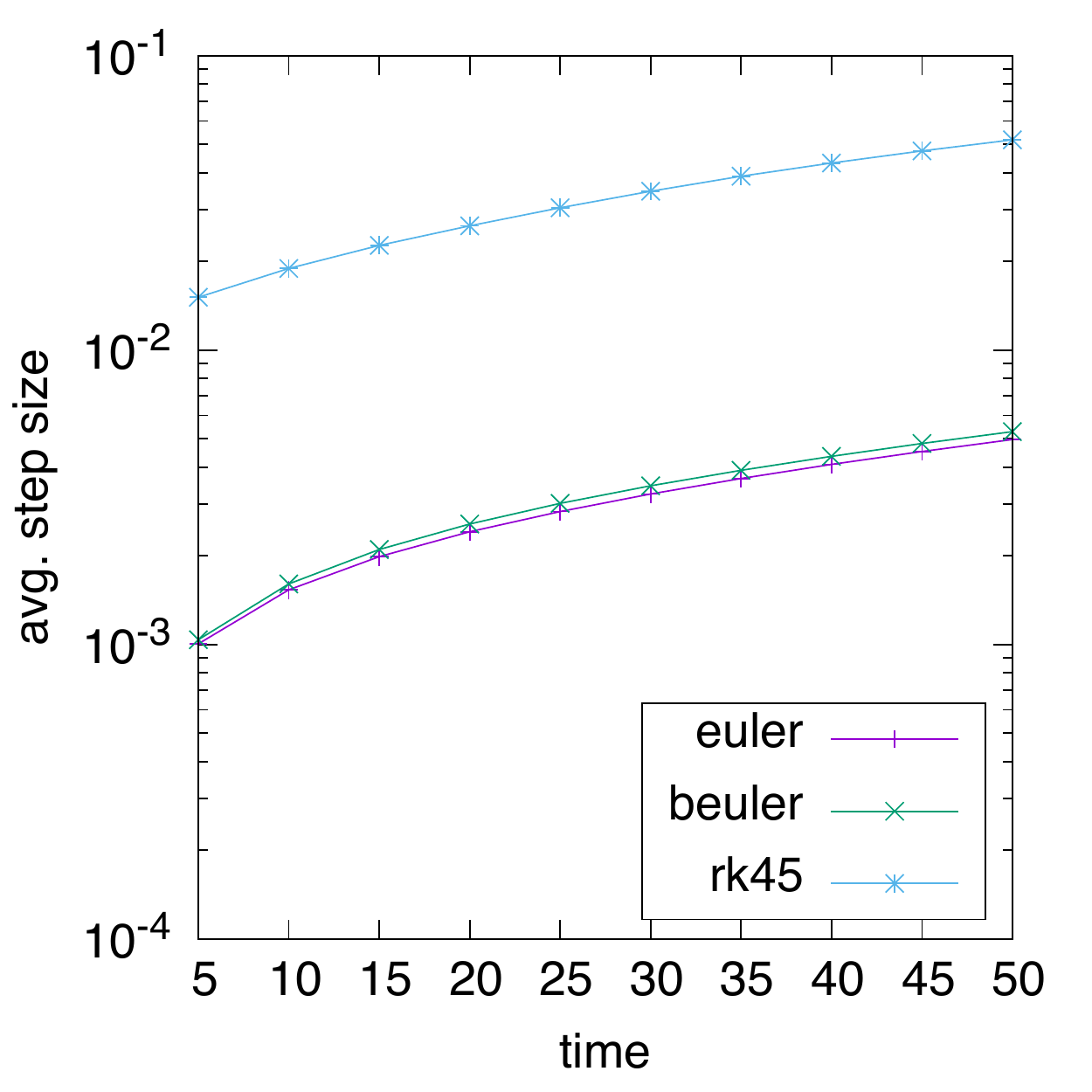}}
\subfloat[][]{\includegraphics[width=0.33\textwidth]{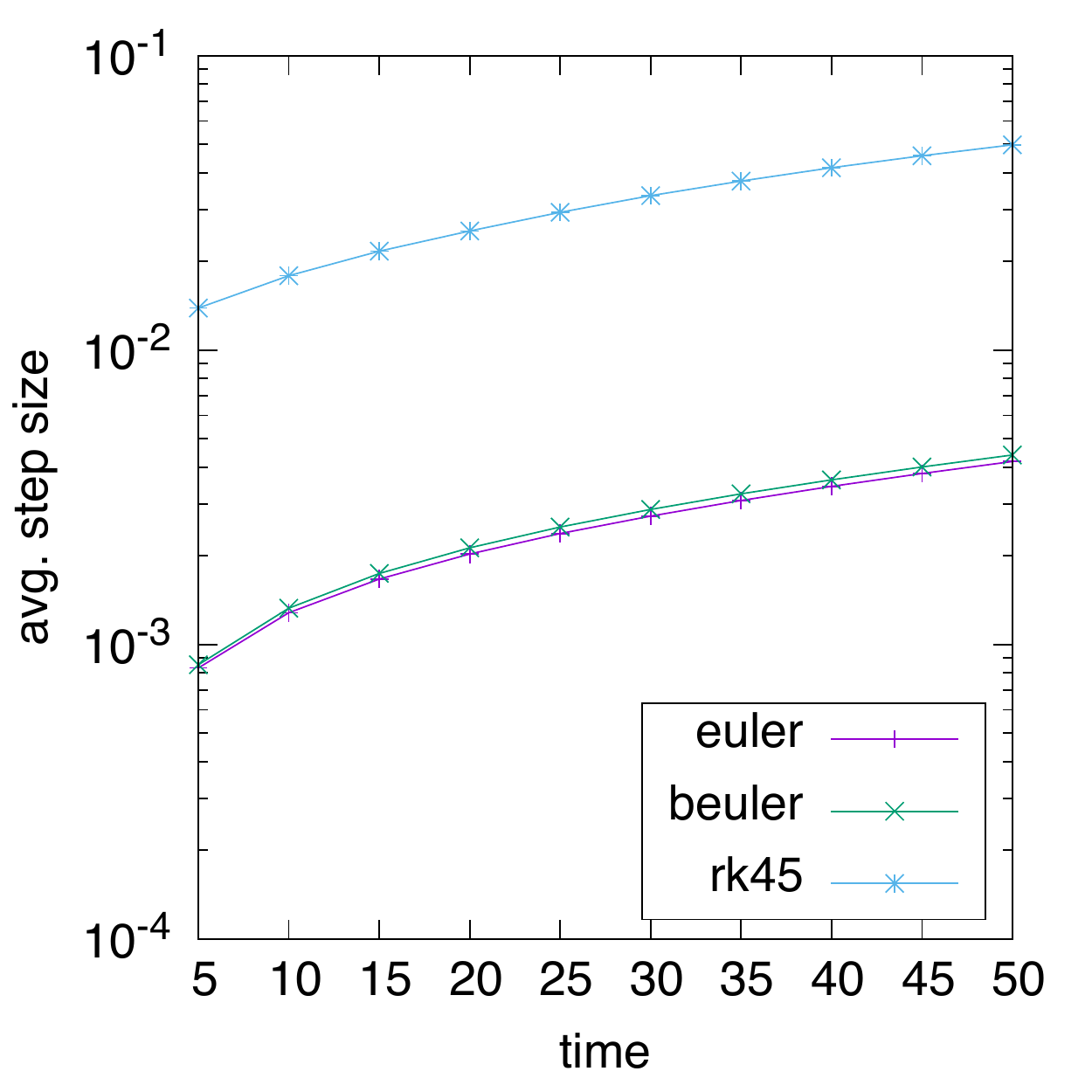}}
\subfloat[][]{\includegraphics[width=0.33\textwidth]{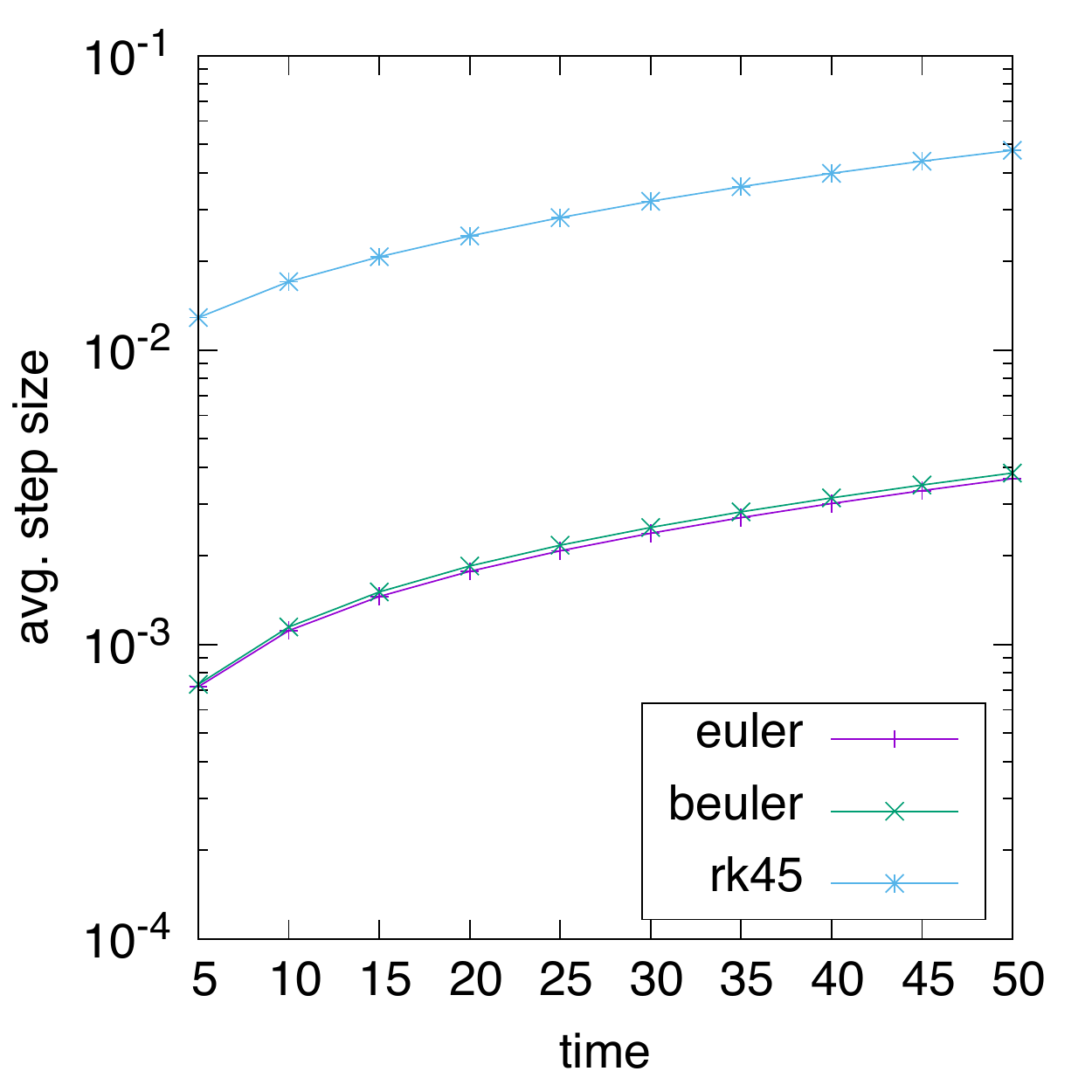}}
\caption{Average step sizes for birth-death process ($\atol=$ (a) $10^{-10}$, (b) $10^{-12}$, (c) $10^{-14}$). }
\label{fig:bd1_stepsize}
\end{figure}

Figure~\ref{fig:bd1_stepsize} shows the average step sizes taken by the different
integration schemes, where \rkfourfive takes much larger steps than the Euler schemes.

\begin{figure}[h]
\centering
\subfloat[][]{\includegraphics[width=0.33\textwidth]{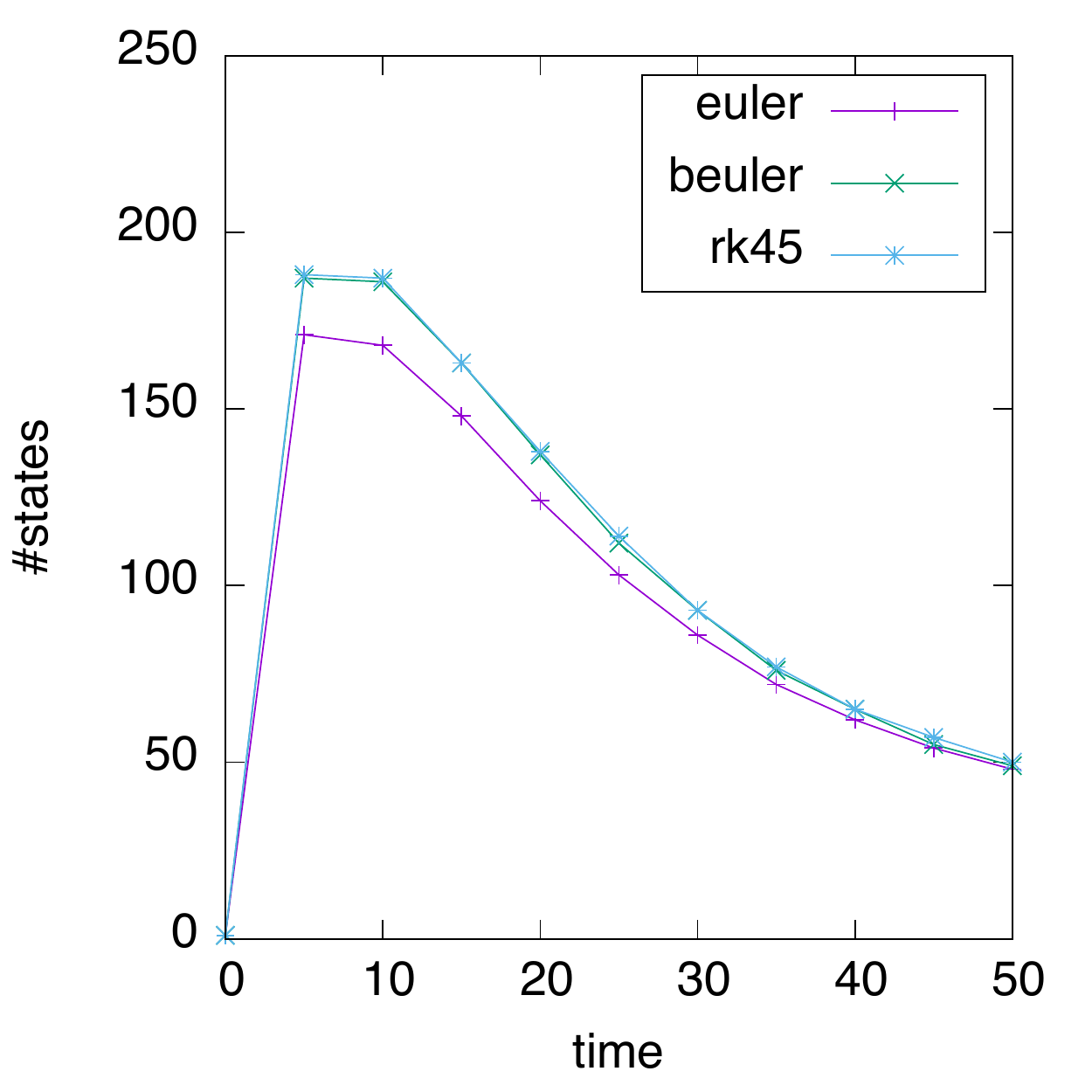}}
\subfloat[][]{\includegraphics[width=0.33\textwidth]{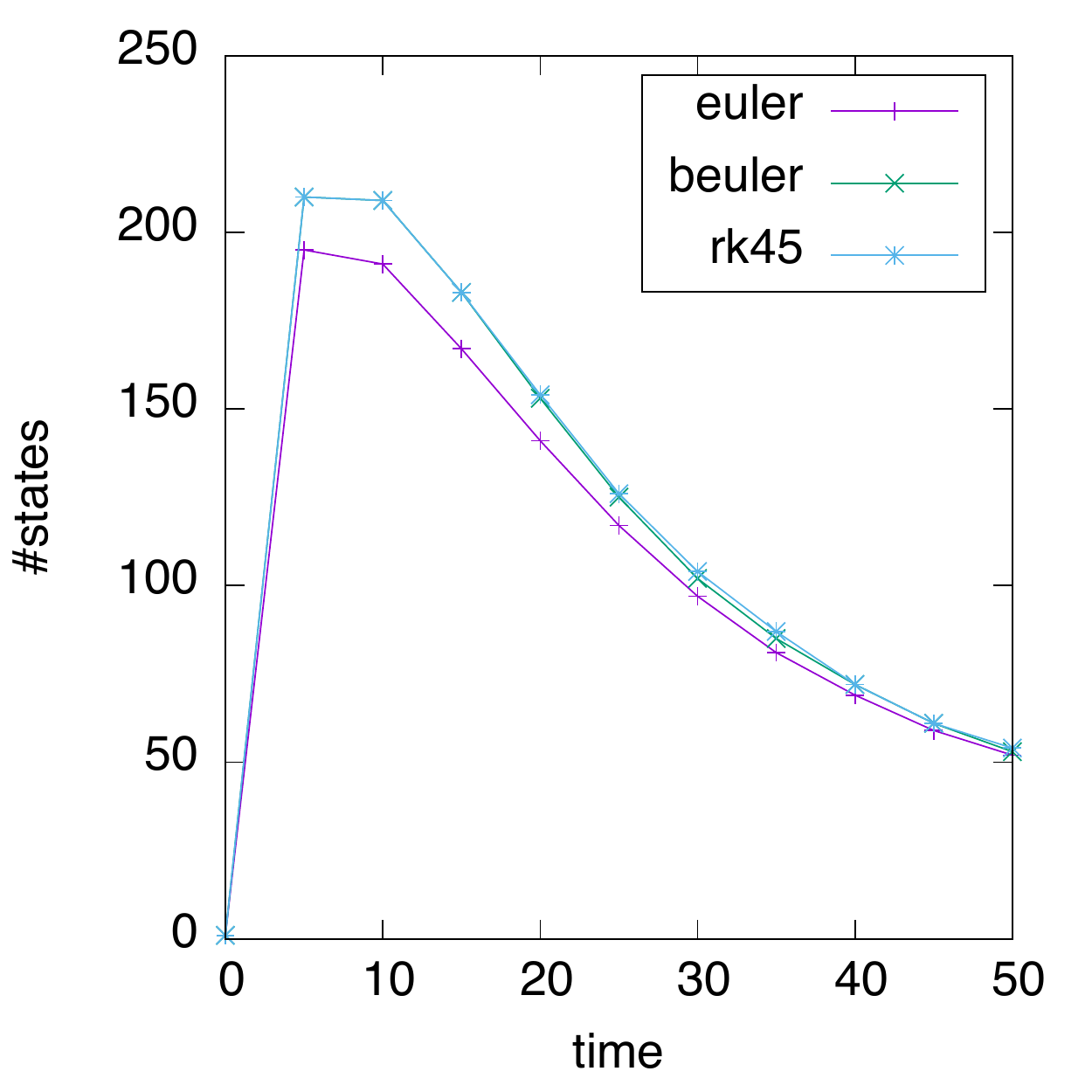}}
\subfloat[][]{\includegraphics[width=0.33\textwidth]{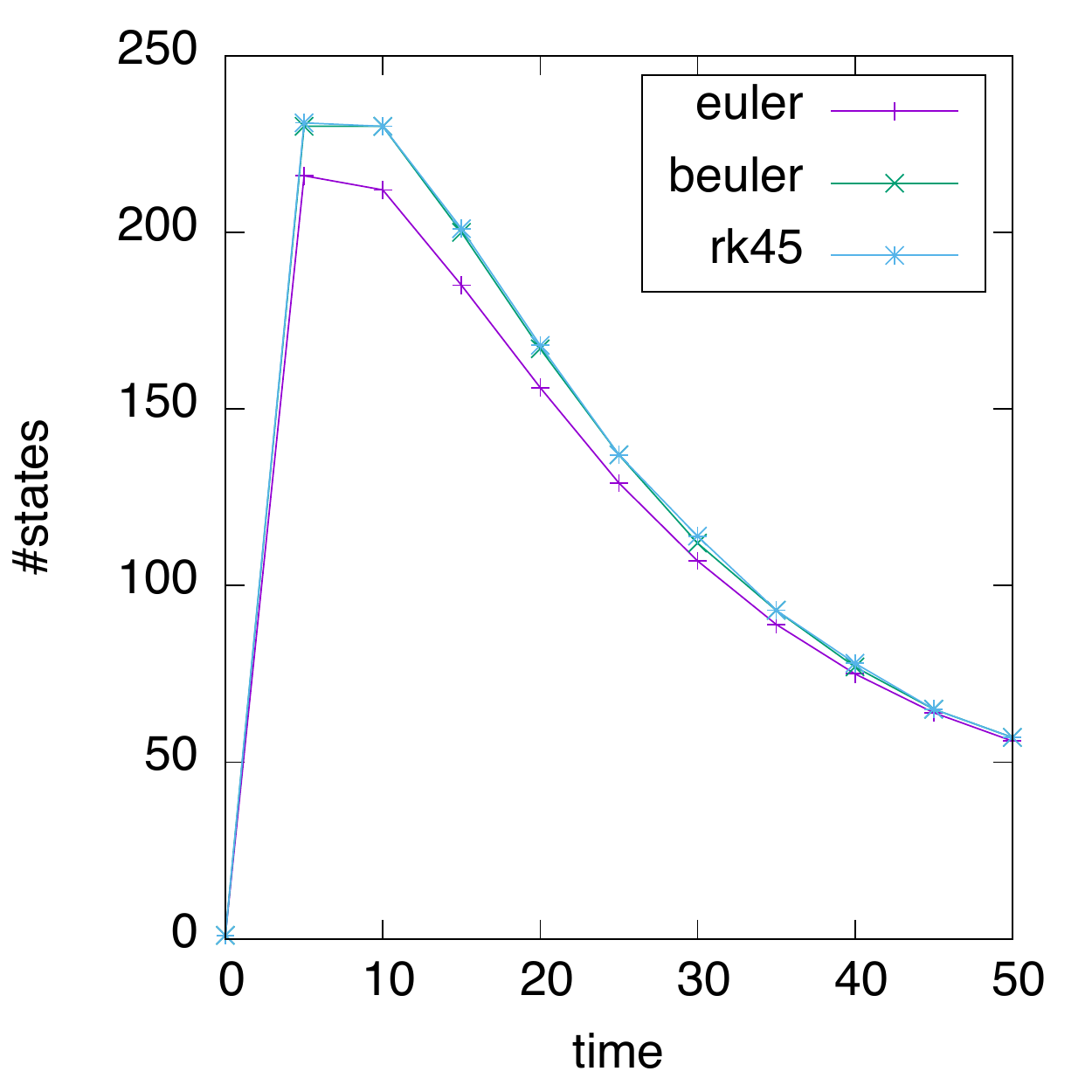}}
\caption{Numbers of significant states for birth-death process ($\atol=$ (a) $10^{-10}$, (b) $10^{-12}$, (c) $10^{-14}$). }
\label{fig:bd1_statespace}
\end{figure}

In Figure~\ref{fig:bd1_statespace} we plot the numbers of significant states used during the computation.
It can be seen that for all considered values of $\atol$ all integration schemes
only require to handle (integrate) a moderate number of states (differential equations)
where the Euler schemes require roughly the same number of states and \rkfourfive
requires only slightly more, in any case for any time less than $250$ states.
Of course, the smaller $\atol$ (and thus our truncation probability threshold $\delta=\atol$)
the larger the number of significant states but with only a slight increase.
This shows that the dynamical truncation procedure indeed substantially reduces
the size of the state space and thus renders possible to integrate numerically
with -- as demonstrated by the previous figures -- maintaining a high accuracy
of the approximations.

\subsection{Yeast cell polarization}
As another reference example we consider a stochastic model of the pheromone-induced
G-protein cycle in the yeast Saccharomyces cerevisiae \cite{chou-etal08,pruyne-bretcher00,roh-etal11}
\begin{alignat*}{6}
\emptyset & \stackrel{c_1}{\longrightarrow}\ R \quad && c_1=0.0038\qquad & RL+G & \stackrel{c_5}{\longrightarrow} G_a+G_{bg} \quad && c_5=0.011\\
R & \stackrel{c_2}{\longrightarrow} \emptyset \quad && c_2=0.0004\qquad & G_a & \stackrel{c_6}{\longrightarrow} G_d \quad && c_6=0.1\\
L+R & \stackrel{c_3}{\longrightarrow} RL+L \quad && c_3=0.042\qquad & G_d+G_{bg} & \stackrel{c_7}{\longrightarrow} G \quad && c_7=1050.0\\
RL & \stackrel{c_4}{\longrightarrow} R \quad && c_4=0.01\qquad & \emptyset & \stackrel{c_8}{\longrightarrow} RL \quad && c_8=3.21\\
\end{alignat*}
with initial state $x(0)=(50,2,0,50,0,0,0)$, where the state vector is given as $x=(R,L,RL,G,G_a,G_{bg},G_d)$
and the state space is the infinite $7$-dimensional set ${\mathbb N}^7$.
Note that in order to keep the meaning of the species here we do not number the species but take
the notation from \cite{roh-etal11}.

\begin{figure}[h]
\centering
\subfloat[][]{\includegraphics[width=0.33\textwidth]{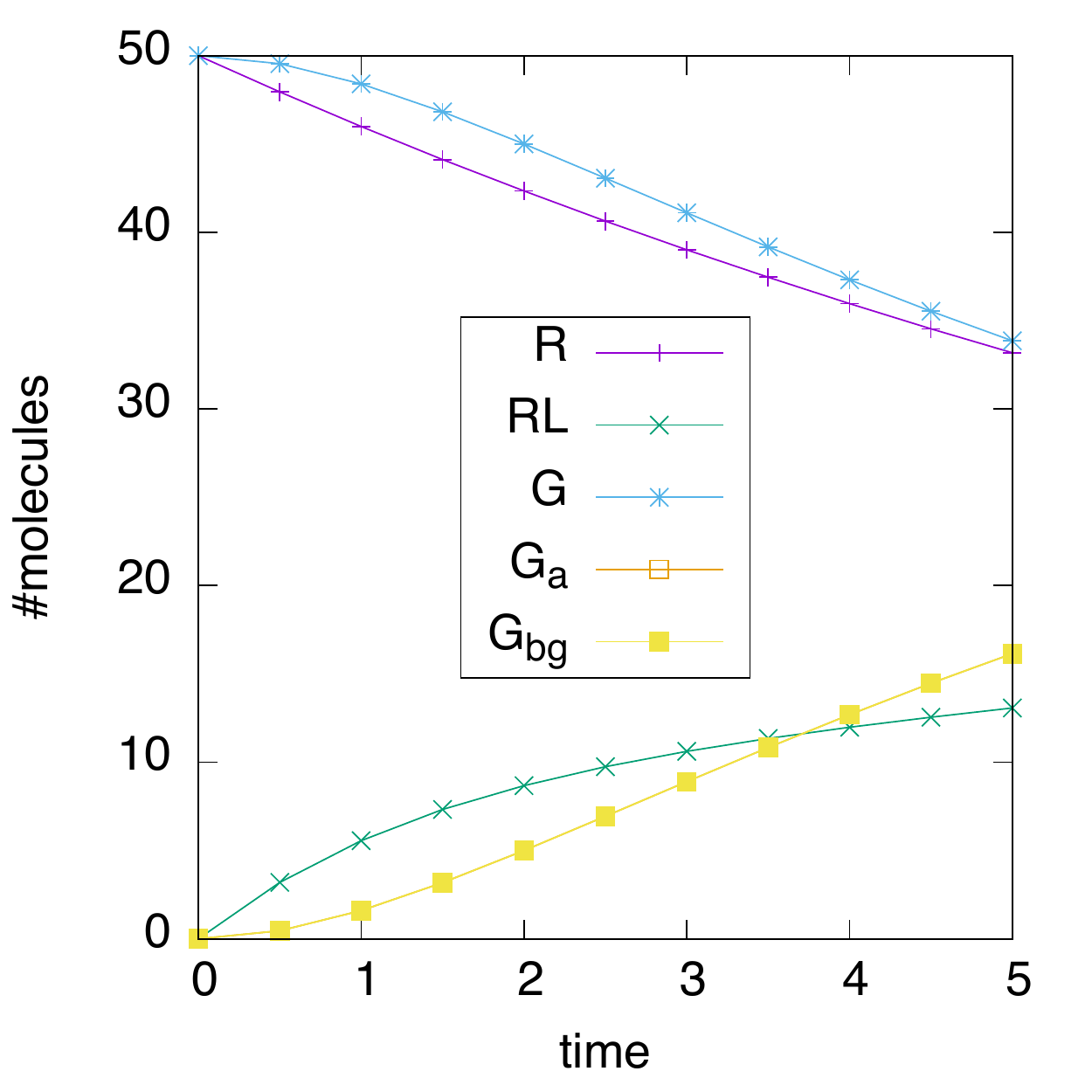}}
\subfloat[][]{\includegraphics[width=0.33\textwidth]{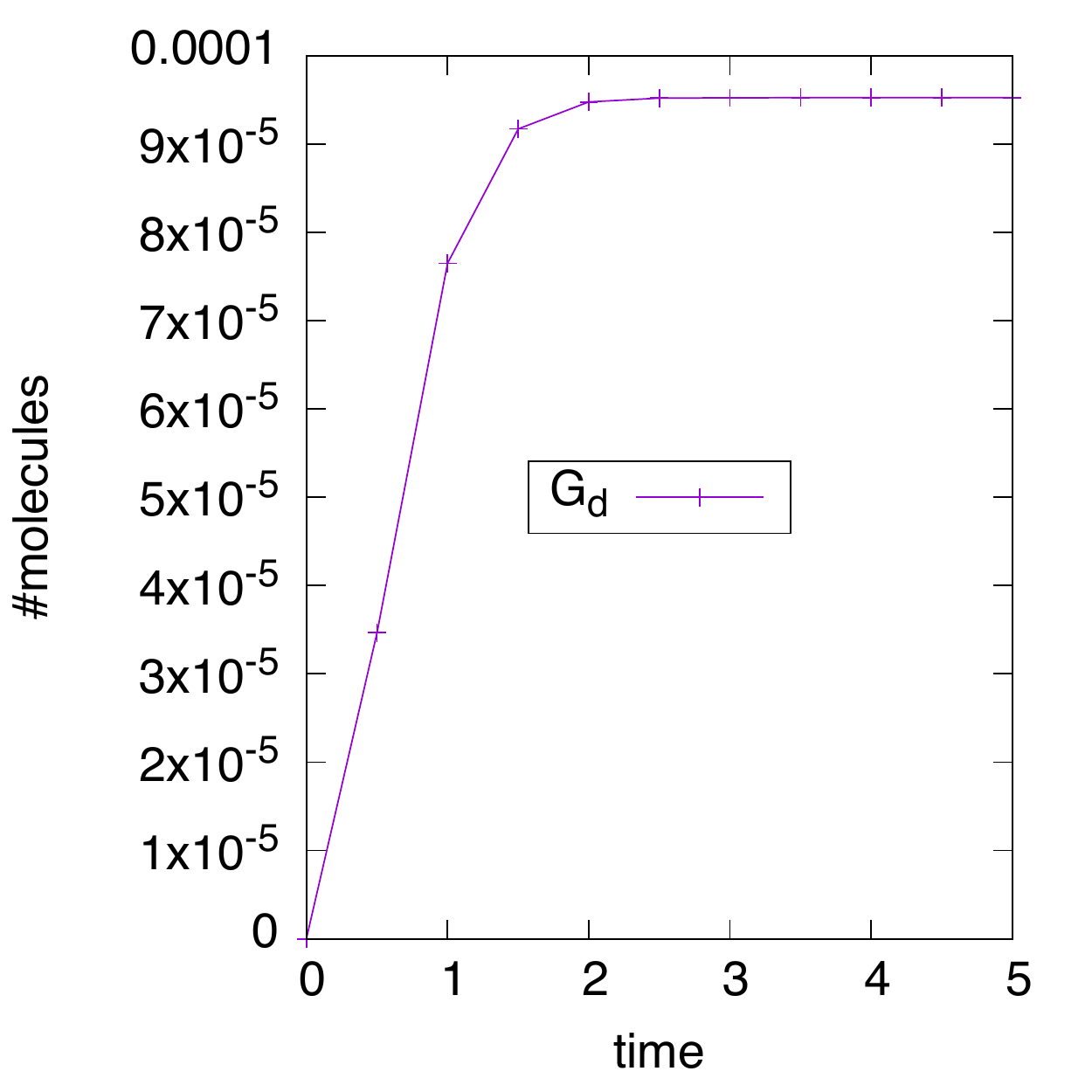}}
\subfloat[][]{\includegraphics[width=0.33\textwidth]{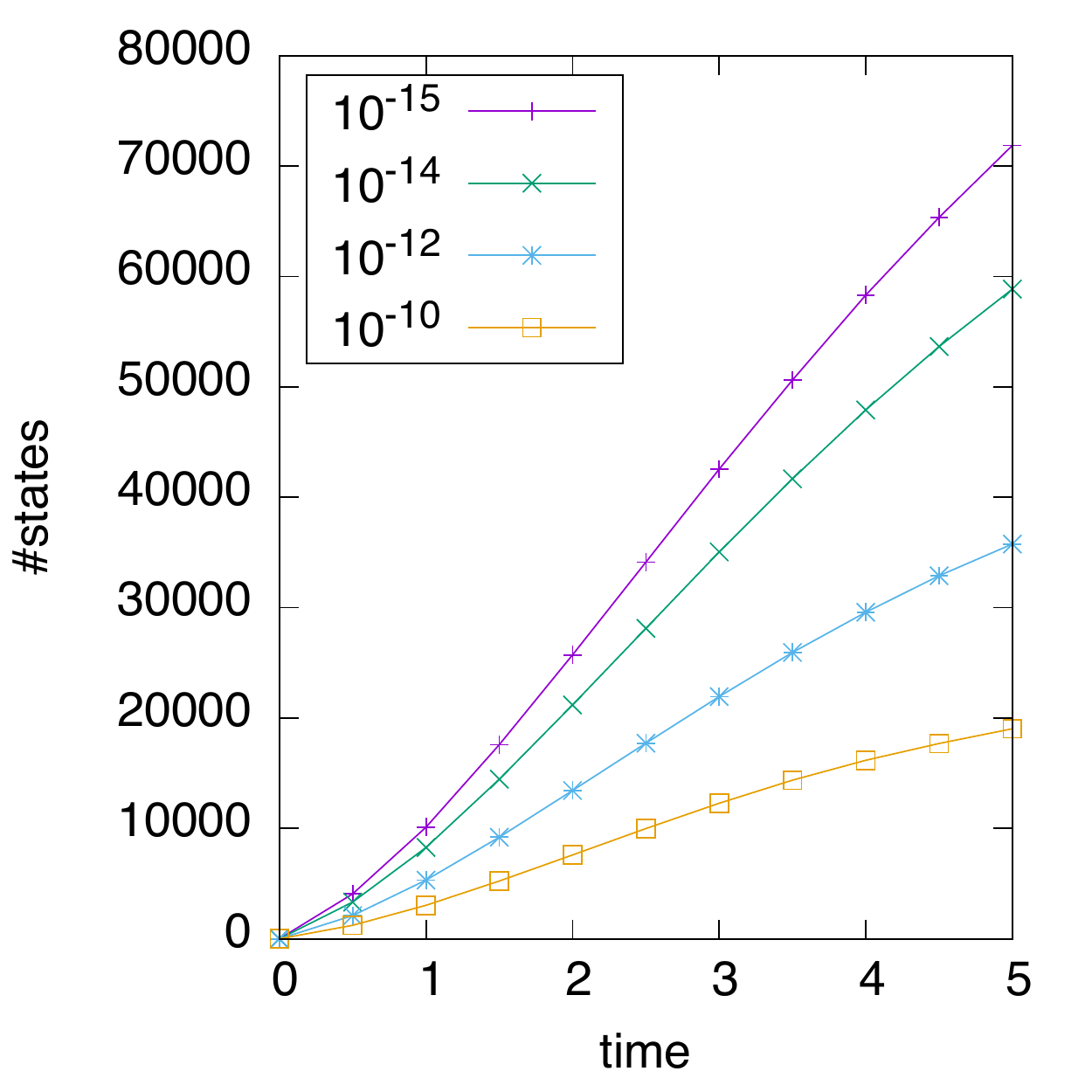}}
\caption{Yeast cell polarization: (a,b) average species counts, (c) number of significant states for different values of $\atol$.}
\label{fig:yeast_means_statespace}
\end{figure}

In Figure~\ref{fig:yeast_means_statespace} we plot the average species counts computed using
\rkfourfive with $\atol=10^{-15}$ and the numbers of significant states for \rkfourfive
over time for
different values of $\atol\in\{10^{-10},10^{-12},10^{-14},10^{-15}\}$.
Note that as in the previous example
the numbers of significant states as well as the average numbers of species
for \euler and \beuler
only slightly differ from those for \rkfourfive, so that we omit to include
them in the plots.
It is clear that in the much more complex yeast cell polarization model there
are many more significant states than in the birth-death process, but the numbers
of significant states are still in a range that allows accurate approximations
in reasonable time.

In Table~\ref{tab:yeast_runtime} we list the run times for different values
of $\atol$. We can see that in this example $\beuler$ heavily outperforms the
explicit methods \euler and \rkfourfive. This confirms the advantages of
implicit methods over explicit methods for stiff systems. In fact, while the
birth-death example is not or only moderately stiff, the yeast cell polarization
model constitutes a very stiff system of differential equations.
\begin{table*}[h]
\caption{Run times for yeast cell polarization model.}
\label{tab:yeast_runtime}
\centering
\begin{tabular}{rccc}
\toprule
method & $\atol=10^{-10}$ & $\atol=10^{-12}$ & $\atol=10^{-14}$ \\
\midrule
\euler & 761s & 1481s & 5839s \\
\beuler & 35s & 76s & 139s \\
\rkfourfive & 5806s & 18603s & 26126s \\
\bottomrule
\end{tabular}
\end{table*}

\begin{figure}[h]
\centering
\subfloat[][]{\includegraphics[width=0.33\textwidth]{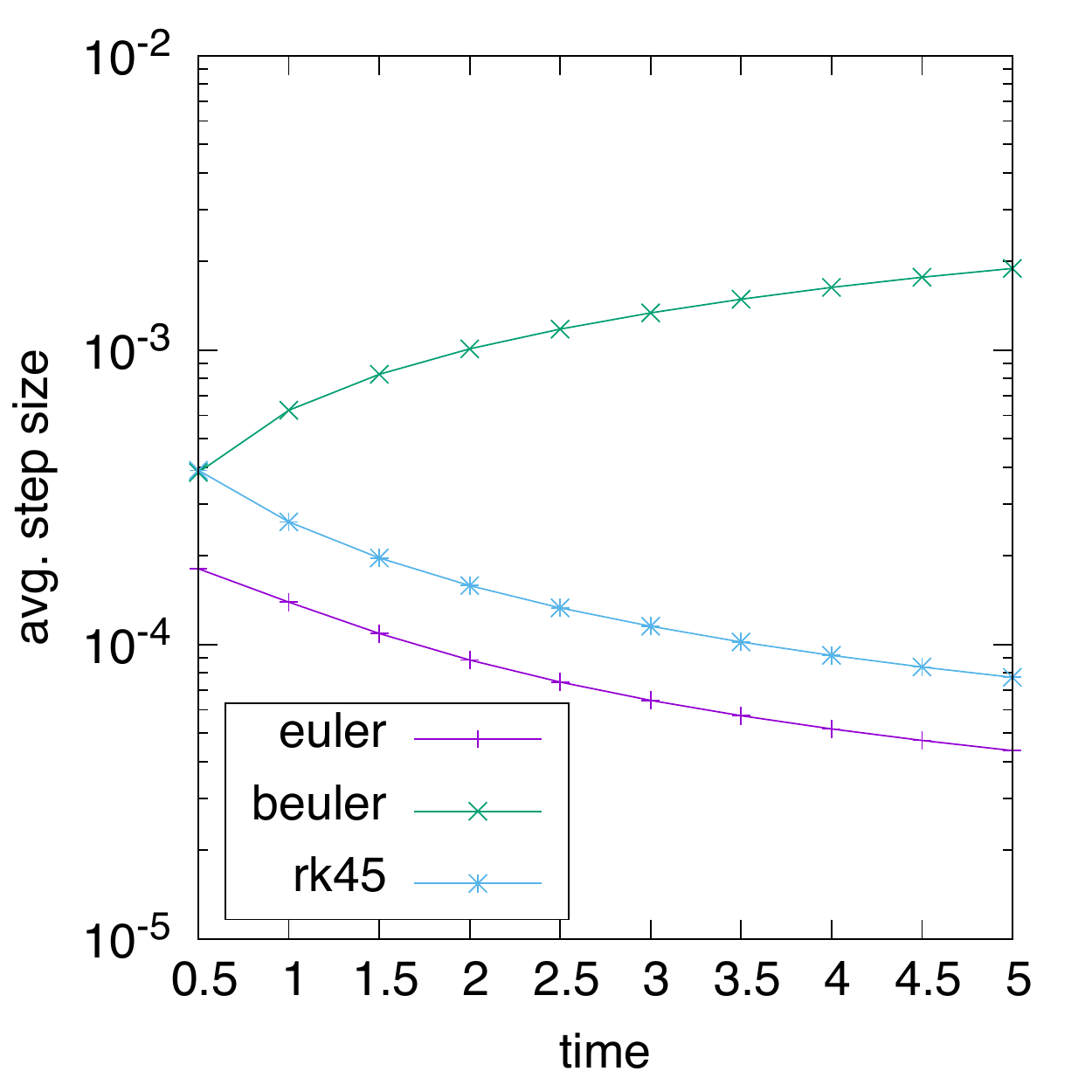}}
\subfloat[][]{\includegraphics[width=0.33\textwidth]{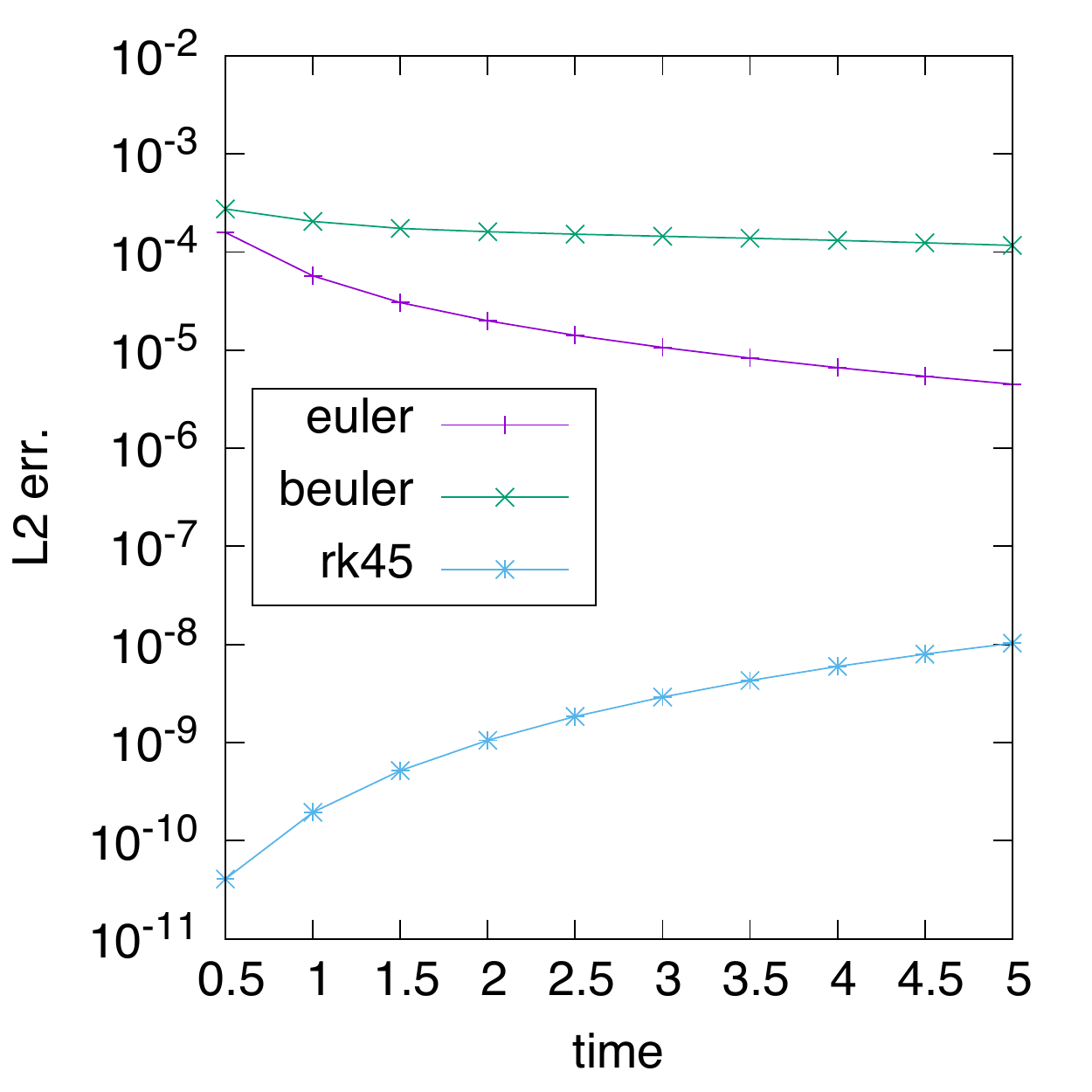}}
\caption{(a) Average step sizes and (b) L2 error for yeast polarization model ($\atol=10^{-14}$).}
\label{fig:yeast_stepsize_err_distr}
\end{figure}

In Figure~\ref{fig:yeast_stepsize_err_distr} we plot the average step size over time and
the L2 error for $\atol=10^{-14}$. Since there is no analytical solution,
we compute the L2 with respect to the distribution obtained using $\rkfourfive$ with
a lower absolute tolerance $\atol=10^{-15}$, applying the rationale that with an
even lower error tolerance the method gives nearly exact values.
The respective L2 errors for $\atol=10^{-10}$ and $\atol=10^{-12}$ are similar.

\begin{figure}[h]
\centering
\subfloat[][]{\includegraphics[width=0.33\textwidth]{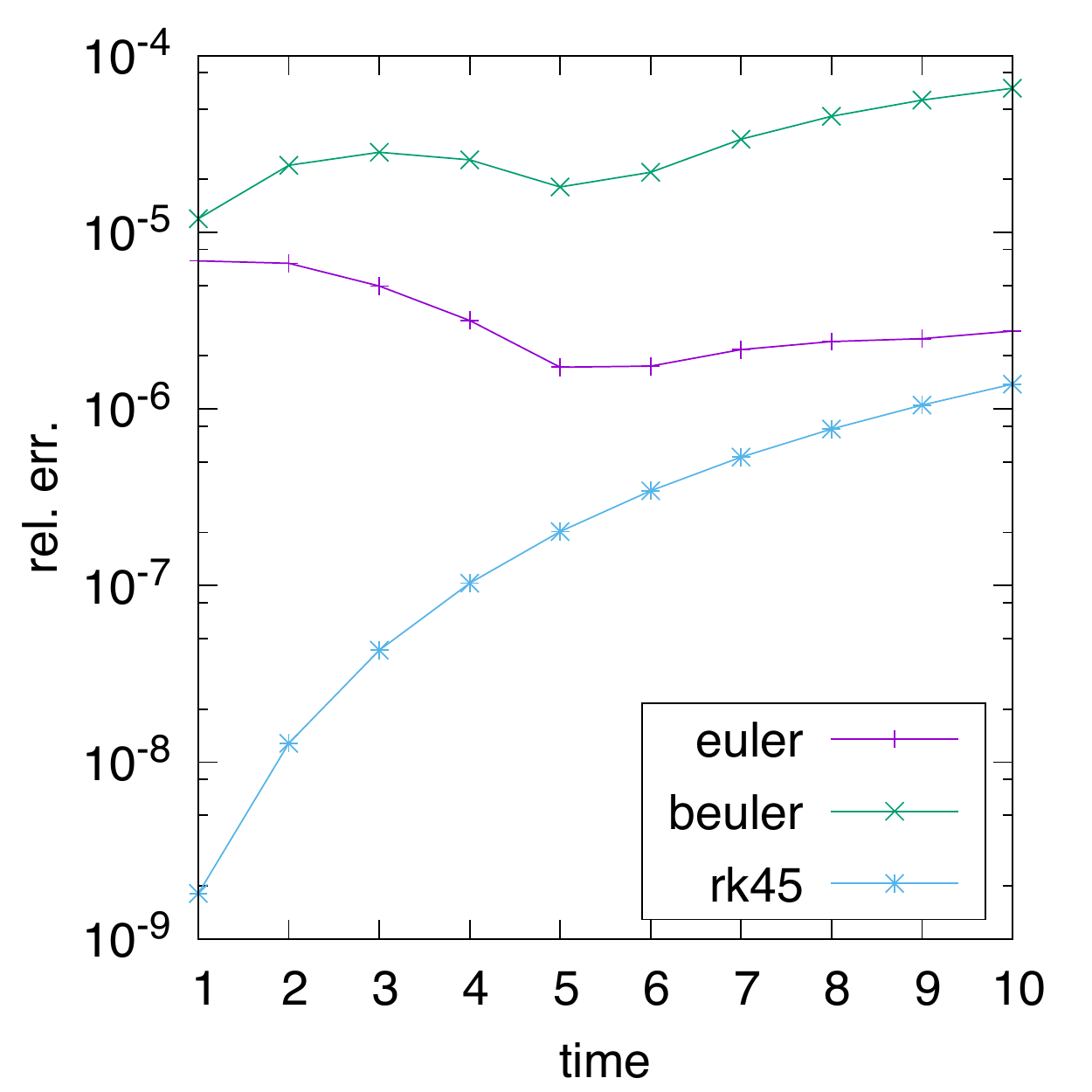}}
\subfloat[][]{\includegraphics[width=0.33\textwidth]{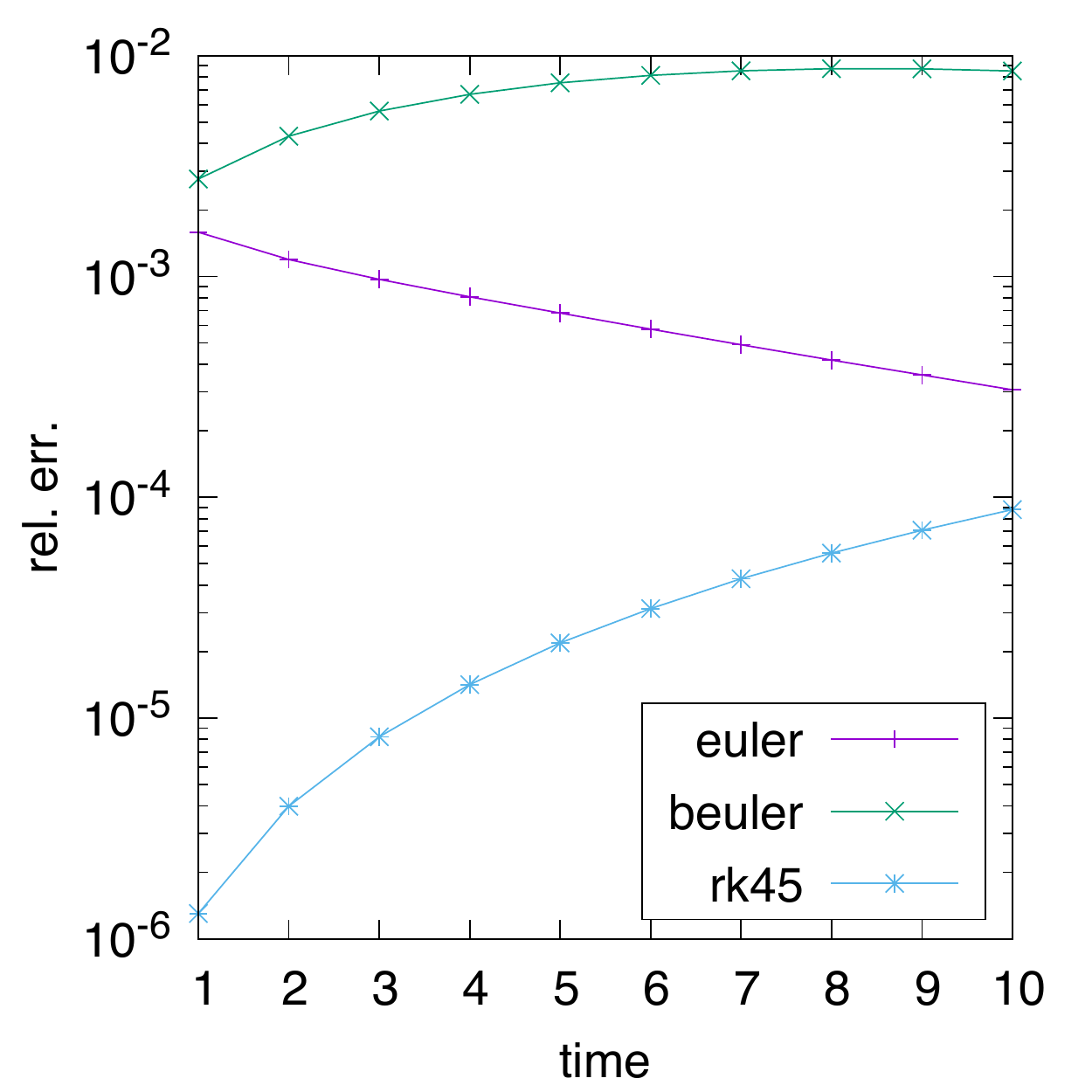}}
\subfloat[][]{\includegraphics[width=0.33\textwidth]{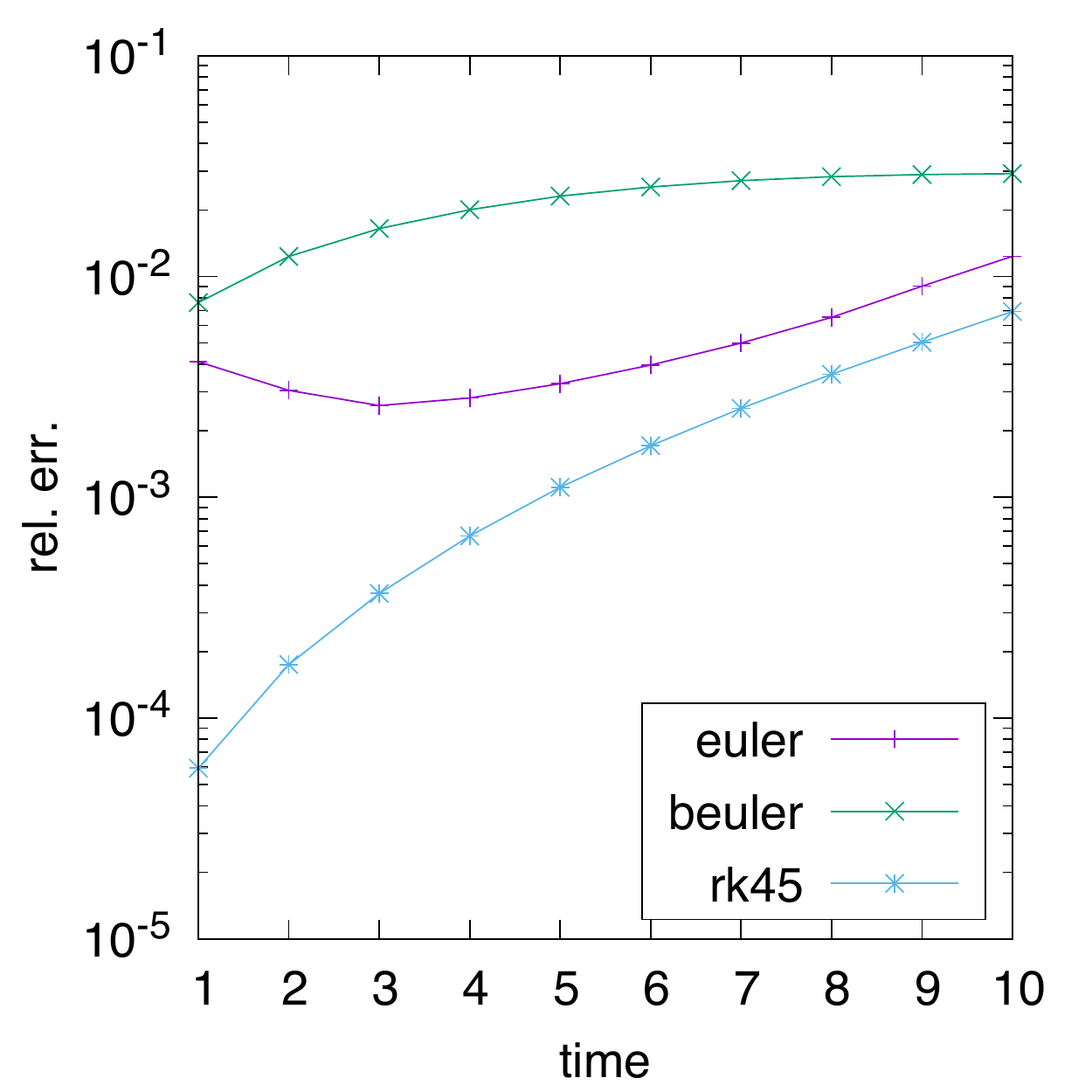}}
\caption{Relative errors.}
\label{fig:yeast_rel_errors}
\end{figure}

\section{Conclusion}
\label{sec:conclusion}
We have shown that our numerical integration approach with adaptive step size selection
based on local error estimates performed in combination with dynamical state space
truncation provides a versatile means of approximating the solution of the chemical
master equation for complex stochastic reaction networks efficiently and accurately.
The state space explosion problem is circumvented by considering in each time step
only such states of the overall state space of the reaction network that have at
that time step a significant (sufficiently large according to a flexibly adjustable
bound) probability, that is, in the course of the integration scheme we keep
the number of differential equations to be integrated per step manageable.
By a framework that includes explicit as well as implicit integration schemes
we offer the flexibility to choose an appropriate integration scheme that is
well-suited with regard to the specific dynamics of a given reaction network.

In order to provide meaningful, detailed comparisons of different methods with
different parameter choices and to study their impact on accuracy, run times
and numbers of significant states to be processed we have considered the explicit
Euler method, an explicit Runge-Kutta as an extension of explicit Euler, and
the implicit (backward) Euler method, all equipped with a well-suited adaptive
step size selection strategy and performed on the dynamically truncated state
space. The results show that the proposed approximate numerical integration of
the chemical master equation indeed yields satisfactorily accurate results
in reasonable time.
Future research will be concerned with further advanced integration schemes,
to equip them similarly with adaptive step size selection strategies and to
study the accuracy and the run times. Of course, also in-depth theoretical
investigations of the performance, efficiency and accuracy of the approximate
numerical integration approach are highly desirable.

\bibliographystyle{plain}
\bibliography{refs-num-ode-markov}

\end{document}